\documentclass[10pt,superscriptaddress,aps,pra,onecolumn,showpacs,floatfix]{revtex4-2}
\usepackage[utf8]{inputenc}
\usepackage{graphicx,amsmath,amsfonts,amssymb}
\usepackage{color}
\usepackage[colorlinks=true, allcolors={blue}]{hyperref}
\usepackage{graphicx}
\usepackage{epstopdf}
\usepackage{float}
\usepackage{placeins}
\usepackage{lipsum}
\usepackage{comment}

\usepackage{soul}

\usepackage{silence}
\usepackage{xcolor}

\usepackage[normalem]{ulem}
\usepackage{orcidlink}

\begin{document}

\preprint{APS/123-QED}

\title{Enhancing Gaussian quantum metrology with position-momentum correlations}

\author{João C. P. Porto \orcidlink{0009-0006-6639-1413}}
\affiliation{Departamento de F\'{i}sica, Universidade Federal do Piau\'{i}, Campus Ministro Petr\^{o}nio Portela, CEP 64049-550, Teresina, PI, Brazil}
\author{Lucas S. Marinho \orcidlink{0000-0002-2923-587X}}
\affiliation{Departamento de F\'{i}sica, Universidade Federal do Piau\'{i}, Campus Ministro Petr\^{o}nio Portela, CEP 64049-550, Teresina, PI, Brazil}

\author{Pedro R. Dieguez \orcidlink{0000-0002-8286-2645}}
\email{Corresponding author: pedro.dieguez@ug.edu.pl}
\affiliation{International Centre for Theory of Quantum Technologies, University of Gdańsk, Jana Bażyńskiego 1A, 80-309 Gdańsk, Poland}

\author{Irismar G. da Paz~\orcidlink{0000-0001-7809-6215}}
\affiliation{Departamento de F\'{i}sica, Universidade Federal do Piau\'{i}, Campus Ministro Petr\^{o}nio Portela, CEP 64049-550, Teresina, PI, Brazil}

\author{Carlos H. S. Vieira~\orcidlink{0000-0001-7809-6215}}
\affiliation{Centro de Ci\^{e}ncias Naturais e Humanas, Universidade Federal do ABC,
Avenida dos Estados 5001, 09210-580 Santo Andr\'e, S\~{a}o Paulo, Brazil.}

\date{\today}

\begin{abstract}
Quantum metrology offers significant improvements in several quantum technologies. In this work, we propose a Gaussian quantum metrology protocol assisted by initial position-momentum correlations (PM). We employ a correlated Gaussian wave packet as a probe to examine the dynamics of Quantum Fisher Information (QFI) and purity based on PM correlations to demonstrate how to estimate the PM correlations and, more importantly, to unlock its potential applications such as a resource to enhance quantum thermometry. In the low-temperature regime, we find an improvement in the thermometry of the surrounding environment when the original system exhibits a non-null initial correlation (correlated Gaussian state). In addition, we explore the connection between the loss of purity and the gain in QFI during the process of estimating the effective environment coupling and its effective temperature. 
\end{abstract}

\maketitle


\section{Introduction}\label{sec:intro}

The search for optimal measurement of properties during classical or quantum noisy processes i.e. determining the ultimate precision in which a parameter can be estimated, after eliminating all technical noise, using ideally accurate instruments, and repeatedly preparing the system in the same state, is crucial for the development of quantum technologies~\cite{Milburnbook}. Quantum metrology offers significant applications such as superresolution imaging~\cite{BraunPRA2023}, high-precision clocks~\cite{RosenbandSCIENCE2008}, estimation of proper times and accelerations in quantum field theory~\cite{Adesso2014SciRep}, navigation devices~\cite{GracePhysRevApplied2020}, magnetometry~\cite{Budker2007NATURE}, optical and gravitational-wave interferometry~\cite{DEMKOWICZDOBRZANSKI2015,TsePRL2019}, and thermometry~\cite{WengPRL2014}. 

Currently, the capacity to measure low temperatures accurately in quantum systems is important in a wide range of proof of principles experiments and has attracted significant interest recently due to its critical role in optimizing the performance of quantum technologies~\cite{DePasquale2018,vieira2023exploring}. 
For instance, effective two-level atoms can be employed to minimize the undesired disturbance on the sample, and an optimum quantum probe (a small controllable quantum system) can act as a thermometer with maximal thermal sensitivity~\cite{AdessoThermometryPRL2015}. The scaling of the temperature estimation precision with the number of quantum probes was investigated~\cite{StacePRA2010}, and the impact of initial system-environment correlations was examined to enhance the estimation of environment parameters within spin-spin model at low temperatures~\cite{mirza2024roleinitialsystemenvironmentcorrelations}. Still in the low-temperature limit, as a fundamental aspect of the third law of thermodynamics,  the unattainability principle of reaching absolute zero also implies the impossibility of precisely measuring temperatures near absolute zero~\cite{BrunnerQuantum2019,Masanes2017NATURE,GallegoPRX2017}. A general approach to low-temperature quantum thermometry considers restrictions from both the sample and the measurement process~\cite{BrunnerQuantum2019}. In continuous-variable systems, a non-Markovian quantum thermometer was proposed to measure the temperature of a quantum reservoir, effectively avoiding error divergence in the low-temperature~\cite{AnJunHongPRApplied2022}. Additionally, optimizing the interaction time in a bipartite Gaussian state allows for precise estimation of the local temperature of trapped ions~\cite{deSaNeto}.

Employing quantum resources such as coherence and entanglement, it is possible to improve, until the Heisenberg limit (HL), the measurement precision beyond the classical shot-noise limit~\cite{Ueda1993PRA,WinelandPRA1994,PlenioPRL2014}. Recently, it was presented a method for low-temperature
measurement that improves thermal range and sensitivity by generating quantum coherence in a thermometer
probe~\cite{Ullah2023PhysRevResearch}.
Optimal quantum metrology strategies usually employ entanglement in initialization and readout stages to significantly enhance measurement precision~\cite{GiovannettiPRL2006,huang2024entanglementenhanced}. However, in practice, producing entangled states or measurements with high precision, in high-dimensional or continuous-variable systems is not a trivial task. Relatively small errors are known to ruin the success of such tasks when compared to the optimal classical one, and the implementation a hybrid quantum-classical approach to automatically optimize the controls has been taken into consideration~\cite{Yang2021SciRep}.

In this work, we employ a
position-momentum (PM) correlated Gaussian state in quantum metrology. To work with it, a real parameter can be controlled such that when it is null, the state recovers the standard uncorrelated Gaussian wave packet form. PM correlations were originally investigated by considering the quantized operators $\hat{x}$ and $\hat{p}$, with $[\hat{x},\hat{p}]=i\hbar$ \cite{bohm1951quantum}. These correlations generalize Glauber's coherent states~\cite{DODONOV1980PLA,Glauber1963} and minimize the Robertson-Schrödinger uncertainty relation. Practically, this parameter arises from atomic beam propagation along a transverse harmonic potential, acting as a thin lens and causing a quadratic phase shift in the initial state~\cite{Janicke1995}. Gaussian correlated packets have applications in various fields, including quantum optics and double-slit matter-wave interferometers~\cite{Campos1999JMO,OzielMPLA2019,LustosaPRA2020,MarinhoPRA2020,Marinho2024SciRep}. However, practical applications face challenges due to errors in tuning the parameter, caused by non-ideal incoherent sources of matter waves and the focalization method used to produce it~\cite{Zeilinger2002,Viale2003PRA,Marinho_2018}. 

%
Here, we propose a scheme to estimate PM correlations.  Our protocol consists of three stages, initialization, interaction, and estimation.
%
Moreover, we further apply our protocol to estimate the effective environment coupling and its temperature (thermometry), and we show that the correlated Gaussian system can outperform the standard (uncorrelated) one. More specifically, we estimated the effective temperature of a Markovian bath via the environmental coupling with our quantum thermometer based on a single-mode correlated Gaussian system. Such correlations impact the Quantum Fisher Information (QFI) after the evolution of the initial state through a Markovian bath and establish the conditions to improve our quantum thermometer model assisted by PM correlations.

The dynamics of the QFI and purity will be analyzed to better estimate this initial position-momentum correlation using the Classical Fisher Information (CFI). Purity, like other measures such as entanglement~\cite{HorodeckiReview2009}, quantum discord~\cite{henderson2001classical,ollivier2001quantum,dieguez2018weak}, and quantum coherence~\cite{Plenio_AdessoRevModPhys2017}, is also a resource, where it is quantified by deviations from the maximally mixed state~\cite{VedralPRL1997,HorodeckiReview2009,ZurekPRL2001,streltsov2014Book,Plenio_AdessoRevModPhys2017,PlenioPRL2014,YangPRL2016,Streltsov_2018NJP}. Purity can be operationally interpreted as the maximum coherence achievable by unitary operations~\cite{Streltsov_2018NJP}. It bounds the maximum amount of entanglement and quantum discord that unitary operations can generate, acting then as a fundamental resource for quantum information processing \cite{Streltsov_2018NJP}. Unlike other measures, purity is easily accessible in experiments, providing experimental bounds to other quantum quantifiers~\cite{HorodeckiPRL2002,ZollerNJP_2013,IslamNATURE2015}. In the following, we present our main results. First, we describe our general estimation protocol in Sec. \ref{sec:result_A}. Then, we investigate the noise environmental effect on PM correlations estimation in Sec. \ref{sec:result_B}. Finally, in Sec. \ref{sec:result_C} we analyze the role of PM correlation in improving the thermal sensitivity in the low-temperature regime, and in Sec.~\ref{sec:disc} we discuss our results.  


\section{Results}\label{sec:results}

    
\subsection{Estimation protocol}\label{sec:result_A}

\begin{figure}[ht]
\centering
\includegraphics[scale = 0.45]{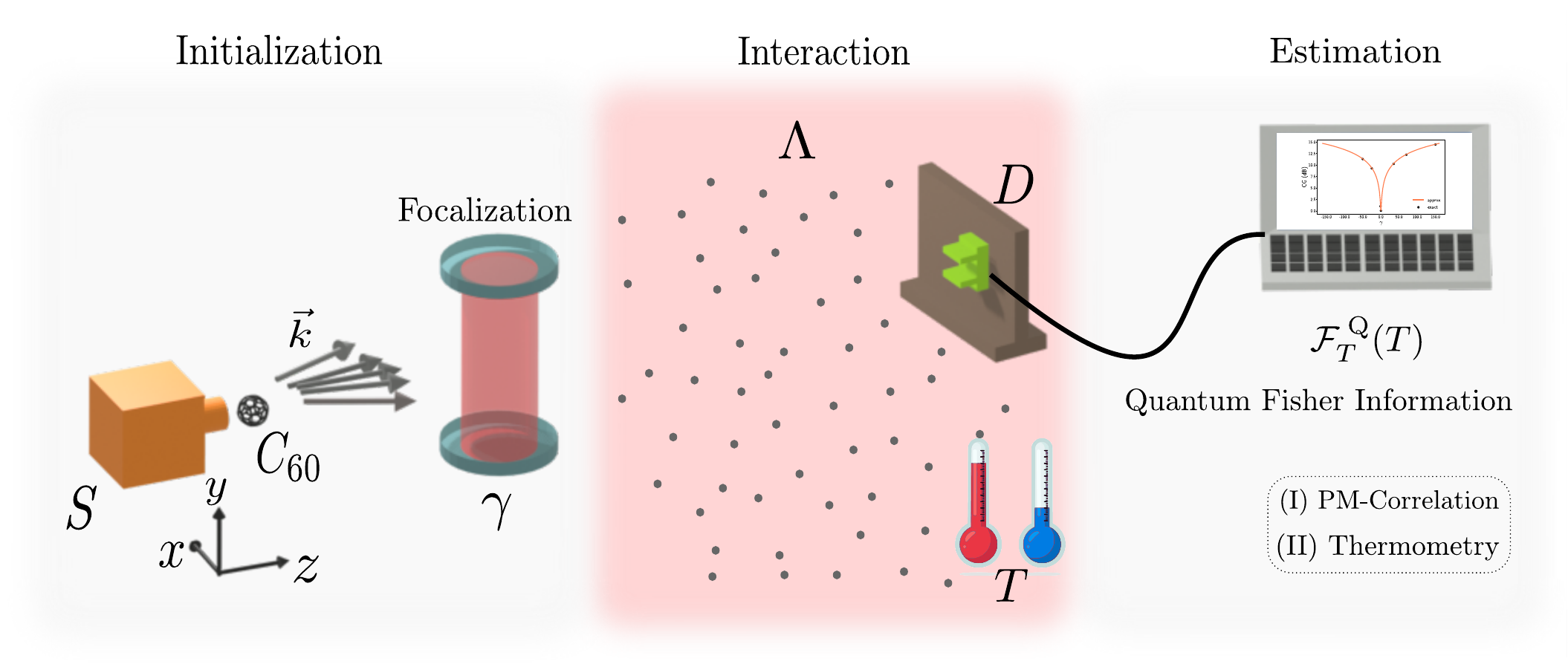}
\caption{Theoretical protocol to describe the estimation of the parameters $\gamma$ (PM correlations) and $\Lambda$ (effective environmental coupling). The initialization step is the source production of fullerene wave packets with momentum $\vec{k}$ and initial coherent length $\ell_0$. Then, the focalization step takes place, generating the initial position-momentum correlations $\gamma $ (see appendix \ref{Ap:correlated_states}). After, the system interacts with the Markovian thermal bath, characterized by the effective environmental coupling $\Lambda$. In the readout stage, parameter estimations are done.}
\label{modelFigure}
\end{figure}

Our quantum estimation protocol schematized in Fig.~\ref{modelFigure}  is represented in three parts: initialization, interaction, and estimation. The initialization step is characterized by the generation of fullerene Gaussian wave packets with momentum $\vec{k}$ and an initial coherent length $\ell_0$. In the sequence, the parameterization process of unknown parameters is separated into two parts. The first one, here denoted as focalization, describes the unitary parameterization associated with the encoding of the PM correlation parameter $\gamma$ employing a real-noisy source (details about this process are discussed in Appendix~\ref{Ap:correlated_states}). The second one, denoted as interaction, represents a non-unitary parameterization for which the probe interacts with a Markovian bath producing an encoding of the effective environment coupling parameter $\Lambda$ and in turn its temperature $T$ in the correlated Gaussian state. Finally, the estimation procedure of the unknown parameters is carried out during the estimation stage. This stage includes (I) PM correlations and (II) thermometry estimations.

We consider as the initial state the following correlated Gaussian state of transverse width $\sigma_{0}$
\begin{equation}\label{psi_0}
\psi_0(x_i)= \frac{1}{\sqrt{\sigma{_0} \sqrt{\pi}}}  \exp
\left[-\frac{{x^2_i}}{{2\sigma^2_0}} + \frac{i \gamma
x^2_i}{2\sigma^2_0}\right], 
\end{equation}
that represents a PM-correlated Gaussian state \cite{dodonov2002nonclassical}. The real parameter $\gamma$ ensures that the initial state is
correlated~
\cite{dodonov2002nonclassical,dodonov2014transmission}. Considering the the initial state, Eq.~(\ref{psi_0}), the variance in position becomes
$\sigma_{xx}=\sigma_{0}^2/2$, whereas the variance in
momentum is $\sigma_{pp}=(1+\gamma^{2})\hbar^2/2\sigma_{0}^2$, and the initial correlations PM is $\sigma_{xp}= \langle ( \hat{x}\hat{p} +\hat{p}\hat{x})/2 \rangle -\langle \hat{x} \rangle \langle \hat{p} \rangle  =\hbar\gamma/2$. Notice that, for
$\gamma=0$ we have a simple uncorrelated Gaussian wave packet, i.e., $\sigma_{xp}=0$. Exploring the Pearson correlation coefficient between $\hat{x}$ and $\hat{p}$, i.e., $r=\sigma_{xp}/\sqrt{\sigma_{xx}\sigma_{pp}}\,(-1\leq r\leq1)$, the $\gamma$ parameter turns out to be $\gamma=r/\sqrt{1-r^{2}}\,(-\infty\leq \gamma\leq\infty)$, illustrating the physical meaning of the $\gamma$ as a parameter that encoded the initial correlations between position and momentum for the initial state. 

In the standard quantum phase estimation protocols, it is usual to estimate the physical parameter that implements a relative phase shift~\cite{HRADIL2005PLA,Lee2012}. Here, our main goal will be to explore whether initial PM correlations can work as a resource to improve quantum metrology under noise. To estimate the initial PM correlations, we consider the production of beam particles from a non-ideal or partially incoherent source (see Fig.~\ref{modelFigure}), where each particle has momentum $p_x=\hbar k_x$ in the $x$-direction within a certain range $\sigma_{k_x}$ of wavenumber $k_x$ and can be modeled by the following initial density matrix \cite{Marinho_2023,Viale2003PRA,PP2023} 
\begin{equation}
\rho_{\gamma}(x,x_0')=\int dk_x p(k_x) \psi_{0k_x}(x_0) \psi_{0k_x}^{*}(x_0'),
\end{equation}
with $\psi_{0k_x}(x_0) =\psi_{0}(x_0)e^{ik_x x_0}$. In this model, we consider wave effects only in $x$-direction since the energy associated with the momentum of the particles in the $z$-direction is high enough such that the momentum component $p_z$ is sharply defined, i.e., $\Delta p_z \ll p_z$. Then, the behavior in the $z$-direction can be considered classical \cite{Viale2003PRA}. The geometry of the collimation device is such that it selects only particles with a transverse wave number $k_x$ inside a specific range defined by the width $\sigma_{kx}$, being described by a classical Maxwell–Boltzmann distribution, $p(k_{x})=e^{-(k_{x}^{2}/2\sigma_{kx}^{2})}/\sqrt{2\pi}\sigma_{kx}$. After some algebraic modifications, the initial density matrix of the coherent correlated Gaussian wave packet becomes
\begin{equation}\label{rho_0}
    \rho_{\gamma}(x_{0},x_{0}')=\mathcal{N}_{0}\exp\left\{ -\mathcal{A}_{0}x_{0}^{2}-\mathcal{B}_{0}x_{0}^{\prime2}+\mathcal{C}_{0}x_{0}x_{0}^{\prime}\right\},
\end{equation}
where $\mathcal{N}_{0}=1/\sigma_{0}\sqrt{\pi}$, $\mathcal{A}_{0}=1/2\ell_{0}^{2}+(1-i\gamma)/2\sigma_{0}^{2} $, $\mathcal{B}_{0}=1/2\ell_{0}^{2}+(1+i\gamma)/2\sigma_{0}^{2}$, $\mathcal{C}_{0}=1/\ell_{0}^2$, and $\ell_0 = \sigma_{kx}^{-1}$ is related with the coherence level of the initial state. It is important to note that, in the limit $\sigma_{kx} \rightarrow 0 $ (ideal collimation), one has $\ell_0 \rightarrow \infty$ ($\mathcal{C}_{0}\rightarrow 0$), meaning that the initial state (\ref{rho_0}) is completely coherent or pure. On the other hand, if $\sigma_{kx} \rightarrow \infty $, then one has $\ell_0 \rightarrow 0 $ ($\mathcal{C}_{0}\rightarrow \infty$), indicating an completely incoherent or mixed state.

According to the protocol depicted in Fig.~\ref{modelFigure}, after the focalization process, the system interacts with an environment that is considered to be a Markovian bath~\cite{PP2024}, and its evolution is given by
\begin{equation}
\rho_{\Lambda}(x,x^{\prime},t)=\int \int dx_0 dx^{\prime}_0 K_{\Lambda}(x,x^{\prime},t;x_0,x^{\prime}_0,0)\rho_{\gamma}(x_0,x^{\prime}_0),
\label{rho_x}
\end{equation}
where 
\begin{gather}
K_{\Lambda}(x,x^{\prime},t;x_0,x^{\prime}_0,0)=\frac{m}{2\pi\hbar t}\;\exp{\Big\{\frac{im}{2\hbar t}\left[(x-x_0)^{2}+(x^{\prime}-x_0)^{2}\right]-\frac{\Lambda t}{3}\left[(x-x^{\prime})^{2}+(x_0-x^{\prime}_{0})^{2}+(x-x^{\prime})(x_0-x^{\prime}_0)\right]\Big\}},\label{propagator}
\end{gather}
is the propagator which includes the environmental effect~\cite{Viale2003PRA,Schlosshauer2}. We consider the propagation of fullerene molecules and the decoherence effect produced by air molecules scattering~\cite{Viale2003PRA}. In this case, the effective scattering constant is given by $\Lambda = (8/3\hbar^2)\sqrt{2\pi m_{\text{air}}}(k_B T)^{3/2}N w^2$, where $N$ is the total number density of the air, $m_{\text{air}}$ is the mass of the air molecule, $w$ is the size of the molecule of the quantum system, $k_B$ is the Boltzmann constant, and $T$ is the bath temperature~\cite{Schlosshauer2}. In the propagator described by Eq.~(\ref{propagator}), it is assumed that the center-of-mass state of the system remains undisturbed by the scattering events, a valid condition for fullerene molecules ($m \gg m_{\text{air}}$)~\cite{Viale2003PRA}. The environmental scattering decoherence is a ubiquitous effect constantly monitoring the position of quantum systems in the cosmos, that can be caused by air molecules, light (optical photons), solar neutrinos, cosmic muons, background radioactivity, and even the universe's 3 K cosmic background radiation \cite{Schlosshauer2}. It is also essential to emphasize that in the derivation of this model, it was also assumed the following conditions \cite{Schlosshauer2,joos2003decoherence}: the system and the surroundings are not initially correlated; when the composite object–environment system is translated, the scattering interaction remains unchanged; the rate of scattering is much faster than the characteristic rate of change in the state of the system; and the distribution of momenta of the particles in the environment is isotropic and obeys a Maxwell–Boltzmann distribution. In physical terms, the $\Lambda$ parameter quantifies the rate at which spatial
coherence over a given distance $\Delta x$ is suppressed, motivating the introduction of a decoherence timescale given by \cite{Schlosshauer2}
\begin{equation}\label{eq:dec_rate}
  \tau_{\text{dec}} =  \frac{1}{\Lambda (\Delta x)^2 }.
\end{equation}
This equation will be useful for interpreting how PM correlation is related to loss of purity and gain of Fisher information. After integration and manipulating Eq. (\ref{rho_x}), we get 
\begin{equation}\label{rho_Lambda}
\rho_{\Lambda}(x,x^{\prime},t)=\mathcal{N}_{t}\exp\left\{ -\mathcal{A}_{t}x^{2}-\mathcal{B}_{t}x^{\prime2}+\mathcal{C}_{t}xx^{\prime}\right\}, 
\end{equation}
where $\mathcal{A}_{t}$, $\mathcal{B}_{t}$, $\mathcal{C}_{t}$ and $\mathcal{N}_{t}$ are parameters that include the interaction with the Markovian bath (see Appendix \ref{appendix:rho_parameters}).

Employing Eq. (\ref{rho_Lambda}), we can cast the purity of the state $\mu(\gamma,\Lambda,t)=\text{tr}\left[\rho^{2}_{\Lambda}(x,x^{\prime},t)\right]$ in the concise form
\begin{equation}\label{eq:purity_new}
\mu(\gamma,\Lambda,t)=\sqrt{\frac{\mathcal{A}_{t}+\mathcal{B}_{t}-\mathcal{C}_{t}}{4\mathcal{A}_{R}\mathcal{B}_{R}+\mathcal{C}_{R}^{2}}},
\end{equation}
where $\mathcal{A}_{R}=\text{Re}[\mathcal{A}_{t}]$, $\mathcal{B}_{R}=\text{Re}[\mathcal{A}_{t}]$ and $\mathcal{C}_{R}=\text{Re}[\mathcal{A}_{t}]$.
This result produces $\mu=1$ when $\ell_0\rightarrow \infty$ ($\mathcal{C}_{t}\rightarrow 0$), i.e., for a completely coherent source and no environment effect $\Lambda=0$. Outside this regime, we always have $\mu(\gamma,\Lambda,t)<1$, typical of a noisy scenario represented by mixed states (details in Appendix \ref{appendix:rho_parameters}). 

In the following subsections, we present 
the dynamics of the QFI and the purity in each part of the parameterization process illustrated in our estimating protocol in Fig. \ref{modelFigure}. We begin estimating PM correlations in a scenery where the quantum system unavoidably interacts with its surrounding environments and then we explore how these correlations can enhance noisy quantum metrology.



\subsection{PM correlations estimation}\label{sec:result_B}

From the mixed Gaussian state described in Eq. (\ref{rho_Lambda}), it is easy to check that the first moments $\vec{d}=(\langle \hat{x}\rangle,\langle \hat{p}\rangle)$ are null and the dimensionless second moments are given by 
\begin{equation}
\sigma_{xx}=\frac{t^{2}}{\tau_{0}^{2}}\left[\frac{1}{2}+2\left(\frac{\tau_{0}}{2t}+\frac{\gamma}{2}\right)^{2}+\frac{\sigma_{0}^{2}}{\ell_{0}^{2}}+\frac{2\Lambda t\sigma_{0}^{2}}{3}\right]; \;\;\;\;\;\;  \sigma_{pp}=\frac{(1+\gamma^{2})}{2}+\frac{\sigma_{0}^{2}}{\ell_{0}^{2}}+2t\Lambda\sigma_{0}^{2}; 
\end{equation}
\begin{gather}
     \sigma_{xp}= \frac{\gamma}{2}+\frac{t}{2\tau_{0}}(1+\gamma^{2})+\frac{t\sigma_{0}^{2}}{\tau_{0}\ell_{0}^{2}}+\frac{t^{2}\Lambda\sigma_{0}^{2}}{\tau_{0}}. 
\end{gather}
Here, $\tau_0 = (m\sigma_0^2)/\hbar$ is the time at which the distance of the order of the wave packet extension is traversed with speed corresponding to the dispersion in velocity \cite{Marinho_2020}. Then, the covariance matrix $\boldsymbol{\sigma}$ and its inverse $\boldsymbol{\sigma}^{-1}$ can be written as

\begin{equation}
\boldsymbol{\sigma}(\gamma,\ell_{0},\Lambda,t)=\left(\begin{array}{cc}
\sigma_{xx} & \sigma_{xp}\\
\sigma_{px} & \sigma_{pp}
\end{array}\right); \;\;\;\;\;\;\; \boldsymbol{\sigma}^{-1}(\gamma,\ell_{0},\Lambda,t)=\mu^2 \text{adj}(\boldsymbol{\sigma}) ,
\end{equation}
where $\sigma_{px}=\sigma_{xp   }$, $\text{adj}(\boldsymbol{\sigma})$ is the adjugate matrix of $\boldsymbol{\sigma}$ and we have used that, for Gaussian states, $\text{det}(\boldsymbol{\sigma})=\mu^{-2}$ \cite{serafini2017quantum}. The mean number of excitations, which is proportional to the energy of the system, can then be expressed for the single-mode system as follows~\cite{Xu2016}:
\begin{equation}\label{Eq_n}
    n = (\sigma_{xx} + \sigma_{pp}+\vec{d}^{2} - 1)/2 =  \frac{1}{2} \Bigg [ \Big(\frac{1}{2}+ \frac{\sigma_0^2}{\ell_0^2} +\frac{2\Lambda \sigma_0^2t}{3} 
 \Big)\frac{t^2}{\tau_0^2} +\frac{\sigma_0^2}{\ell_0^2} +2\Lambda \sigma_0^2t \Bigg] + \Bigg( \frac{t}{\tau_0}\gamma + \frac{3\gamma^2}{4} \Bigg).
\end{equation}
In optics, the quantity $n$ represents the mean photon number. This quantity will be important to analyze how the QFI scales with $n$, allowing a comparison with the shot noise and the Heisenberg limit (HL), more details in Fig.~\ref{Fig_QFIxN}.

Therefore, we can write the QFI for single-mode Gaussian states (see Eq.(\ref{fisher_serafini}) in Appendix~\ref{Ap:FisherInformation}) as following \cite{PinelPRA2013,Liu_JournaLPhysicsA2020}

\begin{equation}\label{eq:QFI_purity}
\mathcal{F}_{\Theta}^{\text{\;Q}}(\Theta)=\frac{\mu^4}{2(1+\mu^{2})}\text{Tr}\{[\text{adj}(\boldsymbol{\sigma})\partial_{\Theta}\boldsymbol{\sigma}]^{2}\}+2\frac{(\partial_{\Theta}\mu)^2}{1-\mu^4}.
\end{equation}
The above expression will be useful for interpreting our results, where we can explicitly understand the role of purity and its derivative as a resource for estimating unknown parameters in the presence of noise. Furthermore, we also investigate the dynamics of Classical Fisher Information (CFI) obtained from the density distribution in position space as
\begin{gather}
\mathcal{F}_{\Theta}^{\text{\;C}}(\Theta)=\int\frac{1}{P(x_{i}|\Theta)}\Bigg(\frac{\partial P(x_{i}|\Theta)}{\partial\Theta}\Bigg)^{2}dx_{i},
\label{eq:CFI_densitymatrix}
\end{gather}
where $P(x_i|\Theta)=\rho_{\Lambda}(x,x^{\prime}=x,t)$. In what follows, the estimated parameters will be the initial position-momentum correlation $\Theta= \gamma$ and the effective environmental coupling constant $\Theta=\Lambda$.


We employ the QFI and CFI for the estimation of the initial correlations. They were calculated from Eqs. (\ref{eq:QFI_purity}) and (\ref{eq:CFI_densitymatrix}) by making $\Theta = \gamma$. After some simplifications, we obtain
\begin{equation}\label{eq:QFIgamma}
\mathcal{F}_{\gamma}^{\;\text{Q}}(\gamma,t,\Lambda)  = \frac{\mu^4}{2(1+\mu^{2})}\; \Phi_{\gamma}(\gamma,t,\Lambda) +2\frac{(\partial_{\gamma}\mu)^2}{1-\mu^4},  
\end{equation}
and
\begin{equation}
\mathcal{F}_{\gamma}^{\;\text{C}}(\gamma,t,\Lambda)  = \frac{1}{8 \sigma_0^{4}B^{4}}\left(\frac{m}{\hbar t} + \frac{\gamma}{\sigma_0^{2}}\right)^{2},
\end{equation}
\begin{figure*}[ht]
\centering
\includegraphics[scale=0.25]{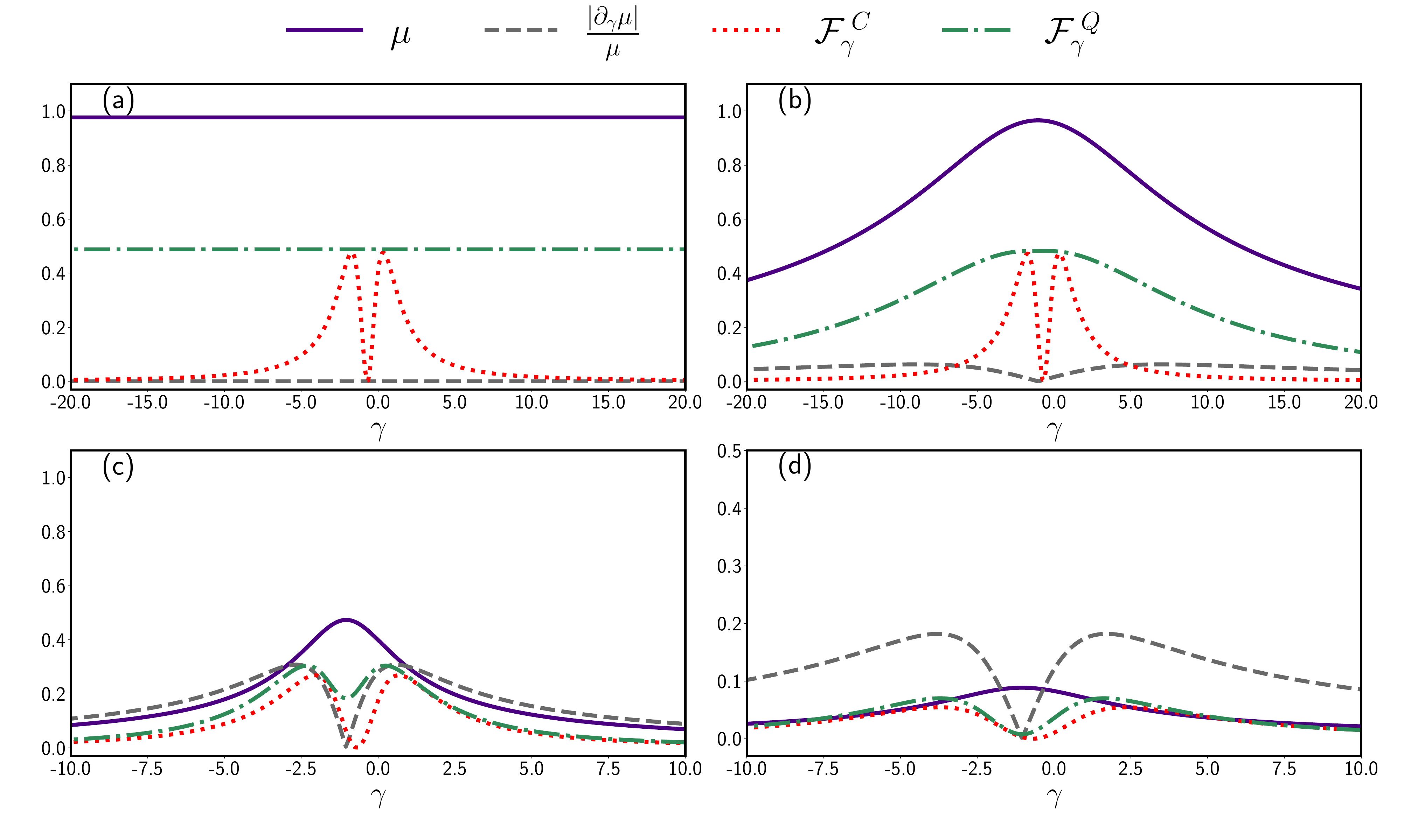}
\caption{Curves for QFI ($\mathcal{F}_{\gamma}^{\text{\;Q}}$), CFI ($\mathcal{F}_{\gamma}^{\text{\;C}}$), purity $\mu$, and its absolute relative derivative $\frac{1}{\mu}|\partial_{\gamma}\mu|$ as a function of the initial correlation $\gamma$ for $t = 1.0 $ $\mu$s, and crescents environmental effect coupling constants: (a) $\Lambda = 0$ ($T=0$ K), (b) $\Lambda = 10^{20}$ m$^{-2}$s$^{-1}$ ($T=952$ K), (c) $\Lambda = 10^{22}$ m$^{-2}$s$^{-1}$ ($T=20.5 \times 10^3$ K), and (d) $\Lambda = 10^{23}$ m$^{-2}$s$^{-1}$ ($T=95.2 \times 10^3$ K). In panels (a) and (b), we observe that when the environment has a small effect (low-temperature regime), the QFI exhibits a behavior similar to the purity. In contrast, panels (c) and (d) show that the QFI exhibits a behavior related to the absolute purity variation when the environment has a strong effect  (high-temperature regime). The advantage of employing PM correlations can be realized for small values of purity.}\label{Fig2}
\end{figure*}
where the parameter $B$ is explicitly defined in Appendix \ref{appendix:rho_parameters} and $\Phi_{\gamma}(\gamma,t,\Lambda)$ is a monotonic increasing function of the parameters $\gamma$, $t$ and $\Lambda$ (see Appendix \ref{Ap:QFIgamma}).  Since this function is multiplied by the fourth power of $\mu$, the purity behavior determines the first term in QFI. This can be illustrated using the following parameters: fullerene mass $m=1.2 \times 10^{-24}$ kg,
molecular size $w = 7$ $\mathring{A}$, width of the initial wave packet $\sigma_0 =7.8$ nm, mass of the air molecule $m_{\text{air}}=5.0 \times 10^{-26}$ kg and density of the air molecules $N = 10^{12}$ molecules/m$^3$  \cite{Marinho_2018,Marinho_2023,Viale2003PRA}. It is assumed that the collimator apparatus selects wave numbers in the $Ox$ direction, with a transverse wavenumber dispersion of $\sigma_{k_{x}} \approx 10^{7}\;\mathrm{m^{-1}}$. This corresponds to an initial coherence length of $\ell_{0} = \sigma_{k_{x}}^{-1} \approx 50\;\mathrm{nm}$. Parameters of this order of magnitude were previously used in experiments with fullerene molecules by Zeilinger \cite{Brezger_PhysRevLett2022,Hornberger_PhysRevLett2003}. With the other parameters ﬁxed, the variation in $\Lambda$ corresponds to the variation in the environmental temperature such that, $\Lambda \propto T^{3/2}$. For instance, $\Lambda$ of the order of $10^{19}$ m$^{-2}$s$^{-1}$, corresponds to an effective temperature around 205 K. The value $3.2\times 10^{15}$ m$^{-2}$s$^{-1}$ for
the scattering by air molecules estimated in \cite{Viale2003PRA} for the
experiment with fullerene molecules corresponds to a temperature of 300 K and a
density of the air molecules of $N = 1.8 \times 10^8$ molecules/m$^3$, smaller than the densities we are considering. In other words, the air molecule density and temperature range we are exploring here are experimentally feasible with the current technology. 

To analyze the role of each term in the QFI (\ref{eq:QFI_purity}) and the influence of the environmental effect, we present in Fig.~\ref{Fig2} the behavior of the QFI ($\mathcal{F}_{\gamma}^{\text{\;Q}}$), CFI ($\mathcal{F}_{\gamma}^{\text{\;C}}$), purity $\mu$, and its absolute relative derivative $\frac{1}{\mu}|\partial_{\gamma}\mu|$ as a function of the initial correlation $\gamma$ for an characteristic interaction time $t = 1.0 $ $\mu$s, and different environmental effect coupling constants.
Panels (a) and (b) demonstrate that the QFI behaves similarly to the purity when the environment has a relatively small effect (low-temperature regime). In contrast, panels (c) and (d) demonstrate that the QFI behaves according to the absolute purity variation when the environment effect is stronger. Also, we see that only in the regime of strong environmental effect (quantum to classical transition in the high-temperature regime) does the CFI become comparable with the QFI. 

One of the most intriguing features here is that besides the suppression of the QFI when the noise increases, we can find PM correlations that enhance the corresponding QFI compared to the usual Gaussian state. It is also important to note that Fig.~\ref{Fig2}(d) shows that even a maximal purity value (corresponding to a null PM correlation) does not provide the maximum value for the QFI. This happens because, in the regime of strong environmental effect, the change in purity (second term in Eq. (\ref{eq:QFIgamma})) is dominant over the purity value itself that is close to zero and will be elevated to the fourth power (first term in Eq. (\ref{eq:QFIgamma})). To comprehend in more detail how the QFI behavior is governed by either the purity or by the purity variation, and how this transition occurs, we present in Appendix~\ref{Ap:QFIgamma} these quantities where we fixed $\Lambda$ and plotted them as a function of the PM correlations and the propagation time.

\subsection{Quantum thermometry assisted by PM correlations}\label{sec:result_C}

\begin{figure}[ht]
\centering
\includegraphics[scale=0.25]{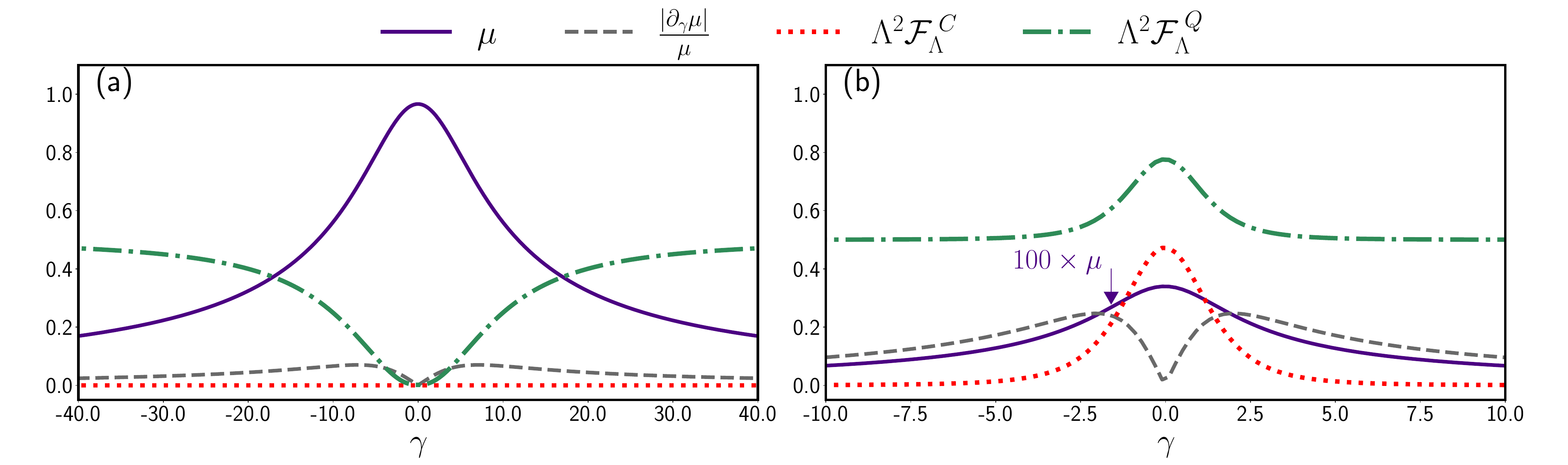}
\caption{Curves for QFI ($\Lambda^2\mathcal{F}_{\Lambda}^{\text{\;Q}}$), 
 CFI ($\Lambda^2\mathcal{F}_{\Lambda}^{\text{\;C}}$), purity $\mu $, and its absolute relative variation $\frac{1}{\mu}|\partial_{\gamma}\mu|$ as a function of initial correlation $\gamma$  for $t = 50.0 $ $\mu$s and considering (a) weak environmental effect ($\Lambda = 1.0 \times 10^{15}$ m$^{-2}$s$^{-1}$ or $T=442$ mK) and (b) strong environmental effect ($\Lambda = 1.0 \times 10^{21}$ m$^{-2}$s$^{-1}$ or $T=4.4 \times 10^3$ K). We observe in panel (b) [(a)] that the gain in Fisher information is related to purity (relative variation in purity) when the environmental effect is strong (weak).}\label{fig3}
\end{figure}

Here, we analyze how the PM correlations affect the estimation of the effective environmental coupling constant, showing a specific regime in which the correlated Gaussian state provides a quantum advantage over the standard one in estimating the effective parameter related to an environmental interaction. 
Using  Eqs. (\ref{eq:QFI_purity}) and (\ref{eq:CFI_densitymatrix}) and setting $\Theta = \Lambda$, the QFI and CFI reads
\begin{equation}\label{eq:QFILambda}
\mathcal{F}_{\Lambda}^{\;\text{Q}}(\gamma,t,\Lambda)  = \frac{\mu^4}{2(1+\mu^{2})}\; \Phi_{\Lambda}(\gamma,t,\Lambda) +2\frac{(\partial_{\Lambda}\mu)^2}{1-\mu^4},  
\end{equation}
and
\begin{equation}
\mathcal{F}_{\Lambda}^{\;\text{C}}(\gamma,\Lambda,t)= \frac{t^{2}}{12 \sigma_0^{4}B^{4}},
\end{equation}
where $\Phi_{\Lambda}(\gamma,t,\Lambda)$ is another monotonic increasing function of the parameters $\gamma$, $t$ and $\Lambda$ (see Appendix \ref{Ap:QFILambda}). Similarly, this function does not determine the profile of QFI in the first term.

In Fig. \ref{fig3}, we exhibit the curves for QFI ($\Lambda^2\mathcal{F}_{\Lambda}^{\text{\;Q}}$), CFI ($\Lambda^2\mathcal{F}_{\Lambda}^{\text{\;C}}$), purity $\mu $, and its absolute relative variation $\frac{1}{\mu}|\partial_{\gamma}\mu|$ as a function of initial correlation $\gamma$ for $t = 50.0 $ $\mu$s and considering (a) weak environmental effect ($\Lambda = 1.0 \times 10^{15}$ m$^{-2}$s$^{-1}$ or $T=442$ mK) and (b) strong environmental effect ($\Lambda = 1.0 \times 10^{21}$ m$^{-2}$s$^{-1}$ or $T=4.4 \times 10^3$ K). Again, we observe two characteristic regimes, the gain in QFI is related either to purity or relative variation in purity when the environmental effect is strong (high-temperature regime) or weak (low-temperature regime), respectively. Such regimes are associated with the first and second term in Eq. (\ref{eq:QFILambda}), respectively. One of the more interesting results of this work, rarely explored in the literature, is investigating the relation between loss of purity and gain in Fisher information. Also, we can observe that in the low-temperature regime, the PM can improve the QFI, and this behavior paves the way for exploring the enhancement of noisy quantum metrology by PM correlations which will be discussed in the following. On the other hand, these PM correlations quantum advantage ceases to be valid in the limit of strong environmental effect ($\Lambda = 1.0 \times 10^{21}$ m$^{-2}$s$^{-1}$ or $T=4.4 \times 10^3$ K), with the standard uncorrelated Gaussian state $\gamma = 0$ providing the best sensitivity for thermometry in this regime [see Fig. \ref{fig3} (b)].

Also, it is important to mention that the effect of increasing the interaction time $t$ or changing the environmental effect $\Lambda$ is almost equivalent from a theoretical point of view, they both decrease the purity and coherence of the quantum state. However, from an experimental point of view, the variation of these two parameters is very different. The first is varied just by moving the position of the detector to regions further away from the source. In contrast, the second can be varied by changing, for example, the environmental temperature or the air pressure.

\begin{figure*}[ht]
\centering
\includegraphics[scale=0.25]{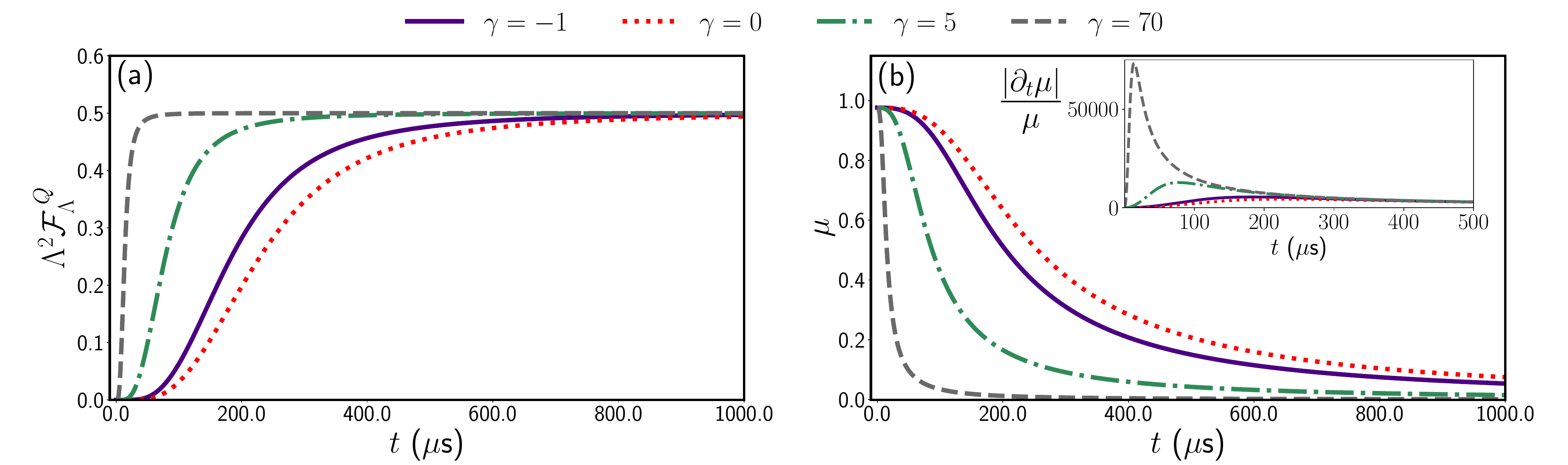}
\caption{ (a) QFI ($\Lambda^2\mathcal{F}_{\Lambda}^{\text{ Q}}$) and (b) purity $\mu $  as a function of time $t$ and for different values of the initial correlation $\gamma$ and considering $\Lambda = 1.0 \times 10^{15}$ m$^{-2}$s$^{-1}$ ($T=442$ mK). We can observe a quantum enhancement due to $\gamma$. Also, note the relation exists between the time of information saturation and the correspondent maximum temporal rate of purity variation [inset of panel(b)].  }
\label{fig4}
\end{figure*}

In Fig.~\ref{fig4} we investigate the time dependence of (a) QFI ($\Lambda^2\mathcal{F}_{\Lambda}^{\text{Q}}$) and (b) purity $\mu $ for different values of the initial correlation $\gamma$ and considering $\Lambda = 1.0 \times 10^{15}$ m$^{-2}$s$^{-1}$ ($T=442$ mK).  We observe a temporal quantum enhancement due to PM correlation $\gamma$. Therefore, employing an initial correlated (position-momentum) Gaussian state can reduce the time required to obtain the maximum amount of information until it is saturated. Moreover, note the relationship [inset of panel (b)] between the maximum temporal rate of purity variations and the time of information saturation. We interpret this gain in QFI as being associated with the rate of change of purity. This relationship can be seen directly through the inset in Fig. \ref{fig4} (b), showing that for large times and regardless of the value of $\gamma$ the second term in Eq.(\ref{eq:QFILambda}) ceases to contribute to an increase in QFI since purity goes to zero at this limit. In other words, the second term contributes to an enhancement in the QFI  only up to approximately  300 microseconds, during which the presence of PM-correlations can further improve this gain. In this context, a similar effect was recently observed in \cite{santos2024improvingparametersestimationgaussian}, where it was noted that the rate of change of coherence with respect to the channel parameter—rather than the amount of coherence—can produce a parameter estimation gain.  
\begin{figure}[ht]
\centering
\includegraphics[scale=0.25]{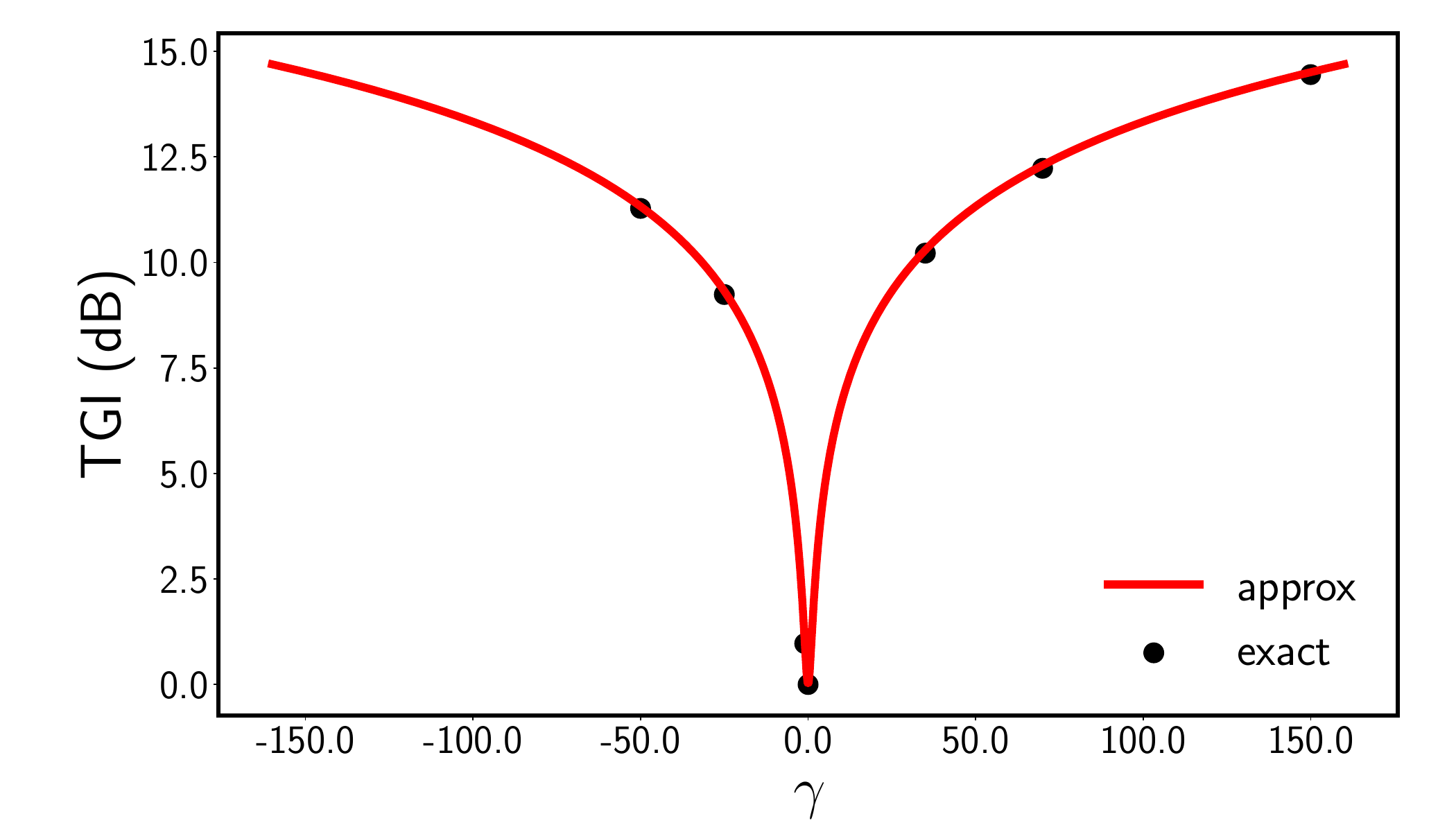}
\caption{Temporal Gain of Information (TGI) as a function of initial PM correlation $\gamma$. The solid red line corresponds to the approximate gain calculated from Eq. (\ref{eq:CGapprox}). The full circles correspond to some values of TGI obtained from the exact value that maximizes the function $\frac{1}{\mu}|\partial_t \mu|$ without any approximation in the purity expression. These points are summarized in Tab. \ref{tab:table1}.}
\label{fig5}
\end{figure}

Besides this, to quantify such an enhancement or acceleration in
the time required to saturate the QFI due to PM correlation $\gamma$, we introduce the Temporal Gain of Information (TGI), defining it on the decibel scale as follows
\begin{equation}
    \text{TGI}(\gamma) =- (10\;\text{dB}) \log  \Bigg [\frac{\tau_{\text{max}}(\Lambda,\gamma)}{\tau_{\text{max}}(\Lambda,\gamma=0)}\Bigg],
\end{equation}
where the time $\tau_{\text{max}}$ corresponds to the maximization point of the relative purity variation, i.e., the maximum of $\frac{1}{\mu}|\partial_t \mu|$ [see inset of Fig. \ref{fig4}(b)]. In this definition, the acceleration in the time to saturate the QFI is compared with the correspondent time needed for the standard uncorrelated Gaussian state ($\gamma = 0$), such that the uncorrelated state is always associated with a null temporal gain.
\begin{table}[b]
\caption{\label{tab:table1}
Values with the time improvement in the estimation environmental effect for $\Lambda=10^{15}$ m$^{-2}$s$^{-1}$.}
\begin{tabular}{|c|c|c|c|c|c|}
\hline
$\gamma$ & $\tau_{\text{max}}$ ($\mu$s)  & $\mu(\tau_{\text{max}})$ & $\frac{1}{\mu}|\frac{d\mu}{dt}|(\tau_{\text{max}})$ $(s^{-1})$ & 
 $\Lambda^2\mathcal{F}_{\Lambda}^{\text{Q}}(\tau_{\text{max}})$ & TGI (dB) \\
\hline
-50.0 & 17.1  & 0.563 & 58 488 & 0.246 & 11.28 \\
-25.0 & 27.2  & 0.563 & 36 861 & 0.247 & 9.24  \\
-1.0 & 183.2  & 0.563 & 5 472 & 0.247 & 0.97  \\
 0.0 & 228.4  & 0.563 & 4 377 & 0.247 & 0 \\
 35.0 & 21.7  & 0.563 & 46 117 & 0.247 & 10.22 \\
 70.0 & 13.7  & 0.563 & 73 191 & 0.248 & 12.23 \\
 150.0 & 8.2  &  0.563 & 121 646  & 0.247 & 14.45 \\
\hline
\end{tabular}
\end{table}

To obtain a simplified and closed expression for the TGI as a function of PM correlations $\gamma$, we consider interaction times in the order of some microsecond and since $(\sigma_{0}/\ell_0)^2 \approx 0.024$ the cubic term in Eq.(\ref{purity}) becomes dominant, and therefore the purity can be approximately expressed as
\begin{gather}
\mu_{\text{approx}}(\gamma,\Lambda,t)\approx\left[1+\frac{4\hbar\Lambda(\gamma^{2}+1)}{3\tau_{0}m}t^{3}\right]^{-\frac{1}{2}}. 
\label{purity_aprox}
\end{gather}
Then, the time $\tau_{\text{max}}$ that maximizes the relative purity variation, i.e., the maximization point of  $\frac{1}{\mu_{\text{approx}}}\partial_t \mu_{\text{approx}}$ is given by
\begin{equation}\label{eq:taumax}
\tau_{\text{max}}(\Lambda,\gamma) = \Bigg[  \frac{3 \tau_0^2}{2(1+\gamma^2)\Lambda \sigma_0^2} \Bigg]^{1/3}.
\end{equation}
Note that this time can be reduced by increasing the initial correlation $\gamma$. In this way, using $\tau_{\text{max}}(\Lambda,\gamma)$ from Eq. (\ref{eq:taumax}), the approximate TGI($\gamma$) can be written explicitly as  
\begin{equation}\label{eq:CGapprox}
\text{TGI}_{\text{approx}}(\gamma)
 =(10\;\text{dB}) \log (1+\gamma^2)^{1/3}.
\end{equation}

In Fig. \ref{fig5}, we examine the enhancement in TGI as a function of PM correlation $\gamma$.  The solid red line corresponds to the approximate gain calculated from Eq. (\ref{eq:CGapprox}). The full circles correspond to particular TGI values determined by a precise number that optimizes the function $\frac{1}{\mu}|\partial_t \mu|$ without any approximation in the purity expression.  These points are obtained directly by probe inspection in the inset of Fig~\ref{fig4}. (b)  and are summarized in Tab. \ref{tab:table1}. Moreover, this procedure was applied to obtain an approximate and simple expression for the TGI and its dependence on the PM correlations $\gamma$. Note that for PM correlation on the order of 150, the gain achievable is almost 15 dB over the standard uncorrelated Gaussian state. In experimental terms, this parameter can be controlled, for example, by changing the laser frequency or intensity in the initialization step (see Appendix \ref{Ap:correlated_states}).

\begin{figure*}[ht]
\centering
\includegraphics[scale=0.3]{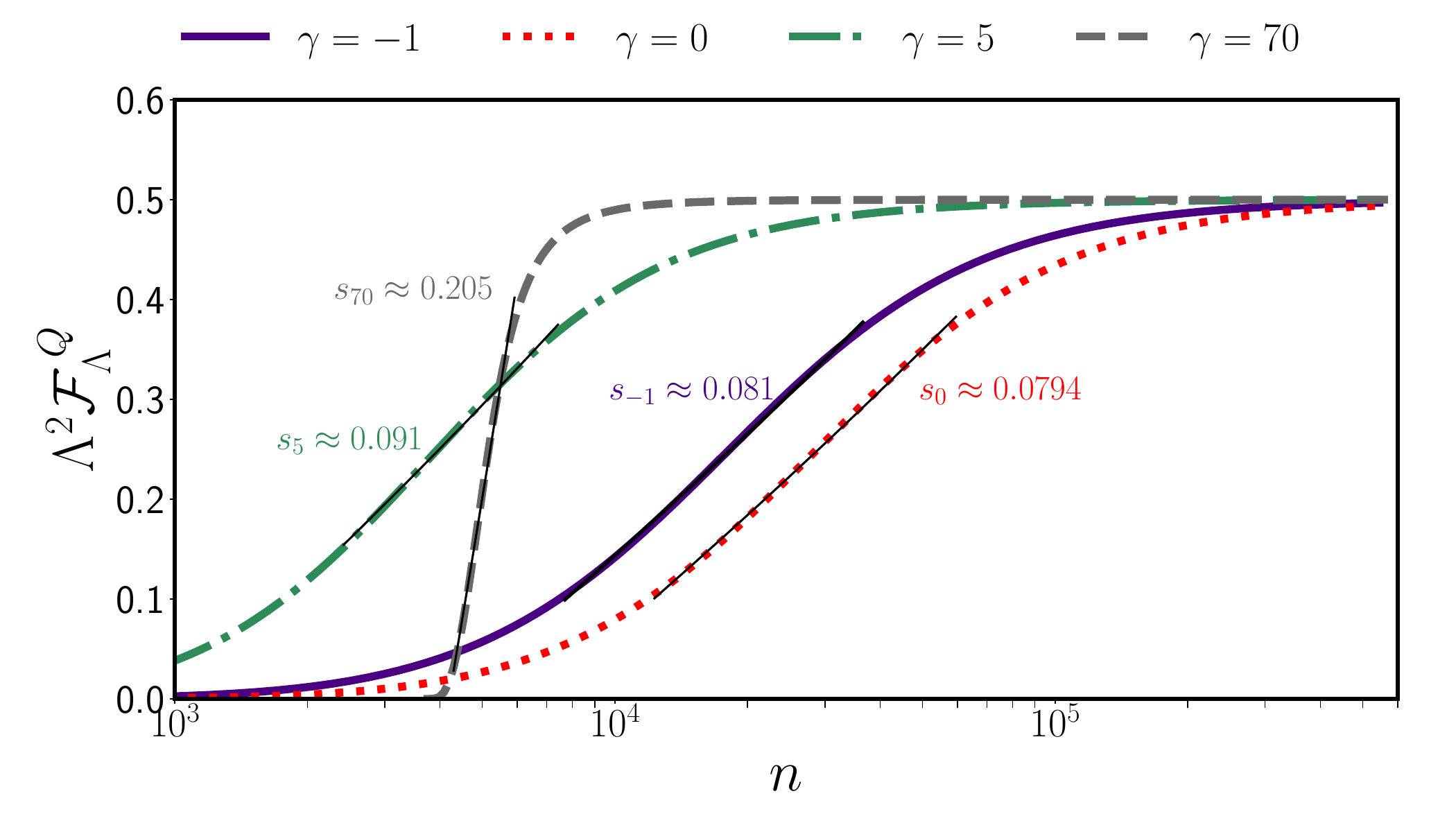}
\caption{ QFI ($\Lambda^2\mathcal{F}_{\Lambda}^{\text{ Q}}$) dependence with the mean number of excitations $n$ in log scale for different values of the initial correlation $\gamma$ and considering $\Lambda = 1.0 \times 10^{15}$ m$^{-2}$s$^{-1}$ ($T=442$ mK). The solid straight lines correspond to a power law fitting of the form $\Lambda^2\mathcal{F}_{\Lambda}^{\text{ Q}} = n^s + b$, where $s$ adjust how the Fisher scales with the mean number of excitations $n$.  We can observe an enhancement in how the QFI scales with $n$ due to $\gamma$.}
\label{Fig_QFIxN}
\end{figure*}

The quantum Cramér-Rao bound often provides a generalized uncertainty relation, even if no Hermitian operator can be simply associated with a given observable \cite{MonrasPRA2011}, as is the case, for phase estimation. In this context, the vacuum fluctuation (shot noise) imposes that one can estimate the phase values with a statistical error $\Delta \Theta$ proportional to $1/\sqrt{n}$, using a classical beam with $n$ average photons or $n$ separate single-photon beams. However, it has been shown that the injection of a squeezed vacuum into the typically unutilized port of an interferometer enables the attainment of a sensitivity of $1/n^{3/4}$~\cite{CavesPRD1981}. Conversely, sensitivity can be further enhanced by employing entangled states, resulting in a resolution that scales inversely with $n$, thereby achieving the Heisenberg limit (HL). To investigate the scaling of the QFI with $n$ and compare the quantum advantage provided by PM correlations against shot noise, squeezing, and the Heisenberg limit, we exhibit in Fig. \ref{Fig_QFIxN}  the QFI ($\Lambda^2\mathcal{F}_{\Lambda}^{\text{ Q}}$) dependence with the mean number of excitations $n$ in log scale for different values of the initial correlation $\gamma$ and considering $\Lambda = 1.0 \times 10^{15}$ m$^{-2}$s$^{-1}$ ($T=442$ mK). The solid straight lines correspond to a power law fitting of the form $\Lambda^2\mathcal{F}_{\Lambda}^{\text{ Q}} = n^s + b$, where $s$ adjusts how the Fisher scales with $n$. This plot was generated in part from Fig. \ref{fig4}, in terms of temporal parameterization $t$ of the QFI behavior and the mean number of excitations $n$ [see Eq. (\ref{Eq_n})]. In the context of statistical error and using the Cramér-Rao inequality (\ref{EQ_CRI}), it can be observed that the sensitivity in estimating the effective environmental coupling constant $\Lambda$ scales as  $1/n^{(1+s)/2}$. The values $s=0$ ($s=1$) correspond to the shot noise limit  (HL), respectively, while $s=1/2$ to the sensitivity of a squeezed vacuum state in a closed quantum system. Thus, the value of $s \approx 0.2$ for $\gamma = 70$ indicates a dependence of $1/n^{0.6}$, demonstrating that PM correlations can be applied to beat the shot noise limit and is almost close to the squeezing dependence of $1/n^{0.75}$ (see also Appendix \ref{Ap_PM_squeezing} where we explore the connection of PM correlation with squeezing), but is still relatively far from reaching HL. Since the generation of entanglement is typically a resource-intensive process, this finding is particularly relevant, as it suggests the potential for metrological enhancement solely through the correlation between two degrees of freedom associated with conjugate observables. This paves the way for advancing quantum metrology without requiring resources like entanglement.

\section{Discussion}\label{sec:disc}

Concepts such as purity distillation and purity cost, analogous to entanglement distillation and entanglement cost, were introduced to quantify this resource~\cite{HorodeckiPRA2003,GourPhysRep2015}. Transformations from pure to mixed states are usually associated with information loss and irreversibility~\cite{HorodeckiPRA2003}. However, this work explores the relationship between Quantum Fisher Information (QFI) and purity rate, focusing on a regime where purity loss primarily drives QFI behavior. It examines how position-momentum (PM) correlated probes can enhance noisy quantum metrology of the environmental thermal bath and demonstrates that PM-correlated Gaussian states can outperform standard uncorrelated Gaussian states in estimating the effective environmental coupling constant.

We started presenting the probe initialization procedure and how the PM correlation can be controlled and produced at this stage. We then calculated the dynamics of QFI and purity for the probe as it evolves through a decoherence channel, representing the scattering model, one of the most common sources of decoherence in physical systems~\cite{Schlosshauer2,joos2003decoherence}. This decoherence model describes how the entanglement of the system with the environmental particles (fermionic or bosonic bath), caused by countless scattering events, delocalizes local phase relations between spatially separated wave-function components, leading to decoherence in position
space (i.e., to localization)~\cite{Schlosshauer2}. Our findings demonstrate the role of purity and purity ``velocity'' $(\partial_{\Theta}\mu)$ as a resource for estimating unknown parameters in the presence of noise [see Eq. (\ref{eq:QFI_purity})].    

We observed that, in estimating the PM correlation, when the decoherence factor is relatively small (i.e., low-temperature regime), the QFI and purity curves have similar behaviors. However, for high temperatures, the QFI is dependent on the rate of change of purity. In this configuration, maximum purity does not necessarily imply maximum values for the QFI due to the dynamics of the purity rate of change term for regimes with an effective strong coupling with the Markovian bath. As expected, QFI tends to match CFI under conditions of quantum-classical transition (high-temperature regime). Furthermore, we found scenarios where the QFI can be asymptotically estimated by the value of CFI by a suitable preparation of the correlated $\gamma$ initial state [see, for example, Fig. \ref{Fig2}(a)]. From a practical standpoint, this result is both surprising and useful, since obtaining the QFI is challenging because it requires maximizing all possible POVMs to ensure that estimations are independent of the chosen POVM and to achieve optimal measurement precision. 

Regarding the estimation of the effective environmental coupling constant, we observed an improvement in the thermometry of the surrounding environment when the probe was prepared in a non-null initial correlated Gaussian state. Specifically, in this part, we explored the acceleration in the loss of purity, due to correlations, as a resource for increasing the QFI [second term in Eq. (\ref{eq:QFILambda})]. Once the parameter $\Lambda$ quantifies the decrease rate in spatial coherence, and since PM correlation values imply that high momentum values are correlated with high spatial values, as a result, we have an extended wavepacket for the probe state. This consequently causes an increase in the decoherence rate $\tau_{\text{dec}}^{-1}$ [see Eq. (\ref{eq:dec_rate})], reducing the purity of the state. 
In our model, we considered the long-wavelength limit \cite{Schlosshauer2}, where the wavelengths of scattering air molecules are much longer than the wave packet extension of the probe. Consequently, a large number of scattering events are needed to resolve the fullerene position by encoding substantial which-path information in the environment, ensuring weak coupling and resulting in an effective Markovian bath. Once we consider that the momentum of environment particles obeys a Maxwell–Boltzmann distribution, the temperature regime investigated here is still more than one million times greater than the ultra-low temperatures obtained, for example, in a Bose-Einstein condensate on the order of 170 nK~\cite{CornellScienceBSC1995}. Similarly, for example, the amplitude-damping channel~\cite{NielsenChuang}, which models physical processes such as spontaneous emission, occurs at $T = 0$ K, and even so, Markovianity is maintained since the coupling of the system with the infinite vacuum modes is weak.

The findings discussed here can be extended to more general scenarios. For instance, recently it was shown that incorporating additional degrees of freedom in molecules, such as rotations and vibrations, can significantly enhance sensitivity when using molecular sensors as probes~\cite{Hutzler_2020QuanScTech,DeMille2024Nature}, creating new opportunities for their application in quantum protocols. Building on this, one could integrate these degrees of freedom into our estimation protocol to fully explore the advantages of molecular probes. Additionally, our findings could also be expanded to include more complex dynamics through a generalized channel that accounts for non-Markovian memory processes. This would offer deeper insights into non-Markovian decoherence features in structured reservoirs, such as decoherence-free states~\cite{FNori2015SciRep}. By pursuing this approach, one could also advance in the exploration at even lower temperatures by relaxing the Born-Markov approximation~\cite{AnJunHongPRApplied2022,BreuerPRA2020}.

In conclusion, the observation that metrological enhancement can be achieved through the correlation of two degrees of freedom associated with conjugate observables is particularly noteworthy, especially considering the typically resource-intensive nature of generating entanglement. This finding paves the way for new explorations in quantum metrology that do not depend on entanglement, indicating a promising direction for future research.

\begin{acknowledgments}
J. C. P. Porto acknowledges Fundação de Amparo à Pesquisa do Estado do Piauí (FAPEPI) for financial support. L.S.M. acknowledges the Federal University of Piau\' i for providing the workspace.  P.R.D acknowledges support from the NCN Poland, ChistEra-2023/05/Y/ST2/00005 under the project Modern Device Independent Cryptography (MoDIC). I.G.P. acknowledges Grant No. 306528/2023-1 from CNPq. C.H.S.V acknowledges the São Paulo Research Foundation (FAPESP), Grant. No. 2023/13362-0, for financial support and the Federal University of ABC (UFABC) to provide the workspace.
\end{acknowledgments}


\appendix

\section{Fisher information}\label{Ap:FisherInformation}

Here, we review the definitions and differences between classical and quantum Fisher information, as well as the definition of purity and its relation to resource theory. Our focus is on describing valid relations specifically for Gaussian states. 

\subsection{Classical Fisher Information}

The Fisher information is a measure of information that an observable random variable carries about an unknown parameter~\cite{Fisher_1925}. It expresses the level of uncertainty of a measured physical quantity. Let $P(x_i|\Theta)$ be the conditioned probability of measuring data $x_i$ given a specific value of $\Theta$. The Classical Fisher Information (CFI) for estimation of the parameter $\Theta$ is defined by 
\begin{gather}
    \mathcal{F}_{\Theta}^{\text{\;C}}(\Theta) = \sum_{i} P(x_i|\Theta)\Bigg(\frac{\partial \ln[P(x_i|\Theta)]}{\partial \Theta}\Bigg)^2 
    =  \sum_{i} \frac{1}{P(x_i|\Theta)} \Bigg( \frac{\partial P(x_i|\Theta)}{\partial \Theta} \Bigg)^2.
\end{gather}

If $P(x_i|\Theta)$ exhibits a peak as a response to variations in $\Theta$, then, the data $x_i$ provides the information needed to estimate the parameter $\Theta$. On the other hand, when $P(x_i|\Theta)$ is flat, numerous samples of $x_i$ would be required to estimate the value of $\Theta$, which would be determined only by using the complete sample population.
In this context, the Cramér-Rao inequality \cite{cramer1946} establishes the lower bound on standard
deviation $\Delta \Theta=\sqrt{\langle \Theta^2 \rangle-\langle \Theta \rangle^2}$ of the estimated parameter 
\begin{equation}\label{EQ_CRI}
   \Delta \Theta \geq \frac{1}{\sqrt{n \mathcal{F}_{\Theta}^{\text{\;C}}(\Theta)}}  ,
\end{equation}
where $n$ is the number of times that the experiment is repeated. Therefore, the Fisher information determines the reachable accuracy of the estimated quantity and represents the figure of merit in parameter estimation problems. Import to mention, that this inequality is valid only for unbiased estimators, i.e., estimators for which $\langle \Theta_{est} \rangle = \Theta_{real}$. 

In quantum theory, we use a set of positive operator-values measure (POVM) $\{\hat{A}(\kappa)\}$, parameterized by $\kappa$, to describe the measurement procedure. These operators are positive and satisfy $\sum_{\kappa} \hat{A}(\kappa)=\hat{I}$ to
ensure normalization. The probability can be expressed as $P(\kappa|\Theta)=\text{Tr}[\hat{\rho}\{\hat{A}(\kappa)\}]$. Using $P(\kappa|\Theta)$ as the probability distribution, the CFI is defined as
\begin{gather}\label{CFI}
    \mathcal{F}_{\Theta}^{\;\text{C}}(\Theta;\hat{A}) = \sum_{\kappa}  \frac{1}{P(\kappa|\Theta)} \Bigg( \frac{\partial P(\kappa|\Theta)}{\partial \Theta} \Bigg)^2,
\end{gather}    
where $\hat{\rho}$ is the density operator describing the system. Naturally, optimal POVMs are those characterized by a statistical distribution of measurement results that is maximally sensitive to changes in the parameter $\Theta$~\cite{Treutlein2018Review}.

\subsection{Quantum Fisher Information}

The Quantum Fisher information (QFI), here denoted by $ \mathcal{F}^{\text{\;Q}}_{\Theta}$, is defined by maximizing Eq. (\ref{CFI}) over all possible POVMs $\{\hat{A}(\kappa)\}$ as follows \cite{Helstrom1969,Holevo_book,Braunstein_CavesPRL1994,Milburn1996}
\begin{equation}
\mathcal{F}_{\Theta}^{\text{\;Q}} (\Theta)  = \max_{\hat{A}} \mathcal{F}_{\Theta}^{\text{\;C}}(\Theta;\hat{A}).
\end{equation}
The quantum estimation is then related to the optimal possible measurement precision. Note that this maximization procedure turns the QFI an upper bound for the classical one, i.e., $\mathcal{F}_{\Theta}^{\;\text{Q}}(\Theta) \geq \mathcal{F}_{\Theta}^{\;\text{C}}(\Theta)$ \cite{Braunstein_CavesPRL1994}. Quantum metrology is not the only application of quantum Fisher information, alternatives include quantum cloning~\cite{SongPRA2013}, entanglement detection~ \cite{SmerziPRL2009,LuoPRA2013}, and quantum phase transition~\cite{ParisPRA2014}. 
Feng and Wei~ \cite{Feng2017} have highlighted the relationship between quantum coherence and Quantum Fisher Information (QFI), demonstrating that QFI is useful for quantifying quantum coherence. This relationship is established because QFI satisfies monotonicity under typical incoherent operations and convexity when quantum states are mixed. In Ref.~\cite{Yadin2021Nature}, it was demonstrated how QFI can be used to identify quantum correlations such as steering, showing that such correlations can be useful for quantum-enhanced measurement protocols. 

For a single-mode Gaussian state, with covariance matrice $\boldsymbol{\sigma}$ and first moments $\boldsymbol{d}$, the QFI is given by \cite{serafini2017quantum,monras2013ARXIV}
\begin{equation}\label{fisher_serafini}
\mathcal{F}_{\Theta}^{\;\text{Q}}(\Theta)=\frac{\text{Tr}[(\boldsymbol{\sigma^{-1}}\partial_{\Theta}\boldsymbol{\sigma})^{2}]}{2(1+\mu^{2})}+2\frac{(\partial_{\Theta}\mu)^2}{1-\mu^4}+2(\partial_{\Theta}\boldsymbol{d})^{\text{T}}(\boldsymbol{\sigma^{-1}})(\partial_{\Theta}\boldsymbol{d}),
\end{equation}
where $\mu = 1/\sqrt{\text{det}(\boldsymbol{\sigma})}$ is the purity of the quantum state and $\partial_{\Theta}$ denotes differentiation with respect to the parameter $\Theta$. The first term is associated with the dynamical dependence of the covariance matrix with the $\Theta$ parameter. The second one is the dynamic of purity under $\Theta$ variation, and the third is the contribution of the moments dynamics of the Gaussian state for the estimated parameter. For example, this result has been utilized in metrological protocols that take advantage of the superradiant phase transition in the Rabi model \cite{FelicettiPRL2020}. For our system under investigation, we will apply the result (\ref{fisher_serafini}) throughout the work to quantify the QFI.

\section{How generate PM-correlated Gaussian states?}\label{Ap:correlated_states}

We review the procedure that can generate correlated Gaussian states for pedagogical reasons and to clarify the work from an experimental point of view. We will explicitly show the parameters determining the initial position-momentum correlation $\gamma$ in Eq. (\ref{psi_0}). The traditional method known as atom lens \cite{Mlynek1992,Mlynek1994} can be applied to produce these initial position-momentum correlations. This method uses the atom-light interactions to create spatially dependent AC-Stark shift
of the electronic ground state of the atoms which is induced in the vicinity of an intensity maximum of a sub-resonant standing wave laser field \cite{Janicke1995}. The standing wave is generated using a cavity perpendicular to the atom direction (see Fig. \ref{modelFigure}). If the atoms are modeled as two-level systems, the atom-laser interaction's Rabi frequency is given by
\begin{equation}
    \Omega (x,z)= \Omega_0 \cos(2\pi x/\lambda) e^{-\pi z^2/(V_{\text{CM}}t_{\text{int}})^2}
\end{equation}
to describe a standing wave with period $\lambda$ (light wavelength) in the $x$-direction and a Gaussian beam profile in the $z$-direction. Here, $V_{\text{CM}}$ is the atomic center-of-mass velocity, and $t_{\text{int}}$ is the effective interaction time, assumed smaller than the radiative lifetime of the excited state of the atoms so that spontaneous emission can be neglected. This approach also applies to the focusing of molecules, since we can consider their vibrational states practically degenerate as long as the energy difference between them is much smaller than the excitation energy of the light beam inside the optical cavity. Due to the Stark shift, the optical potential for incoming atoms in the ground state is given by
\begin{equation}
    U(x,z) =  -\frac{1}{2} \sqrt{\Omega (x,z)^2+\delta^2},
\end{equation}
where $\delta = \omega - \omega_0$, with $\omega = 2\pi c/\lambda$ is the laser frequency and $\omega_0$ is the resonance transition \cite{Janicke1995}. In the vicinity of a maximum intensity region, the atoms roughly feel the harmonic potential $ U(x)\approx \frac{1}{4}\Omega_0 k^2 x^2 (1+\delta^2/\Omega_0^2)$, with photon momentum $k = 2\pi / \lambda$. For a short interaction time, the harmonic potential changes the initial state to
\begin{equation}
    \psi_0(x) \rightarrow  e^{\frac{imV_{\text{CM}}x^2}{2\hbar R(z)} }\psi_0(x),
\end{equation}
where $ R(z)$ is the curvature radius of the wavefronts associated with the beam propagation. Then, it acts as a thin lens with focal length \cite{Janicke1995}
\begin{equation}
    f = \frac{\lambda^2}{\pi \Omega_0 t_{int} \lambda_{\text{dB}}} \Bigg(1+\frac{\delta^2}{\Omega_0^2} 
 \Bigg)^{1/2},
\end{equation}
where $\lambda_{\text{dB}}= h/(mV_{\text{CM}})$ is the de Broglie wavelength of the atoms. We see therefore from this analysis which experimental parameters determine the initial position correlation characterized by the $\gamma$ parameter encoded in the quadratic phase of the initial state (\ref{psi_0}). From an experimental point of view, the focal length $f$ can be varied, for example, by modulating the laser power. Also, note that positive (negative) values of $\gamma$ are associated with a diverging (converging) beam with curvature radius $R > 0$ ($R < 0$).

\section{Parameters of density matrix for a correlated Gaussian state and purity}\label{appendix:rho_parameters}

Here, we describe the parameters obtained through the action of the time evolution propagator over the density matrix shown in Eq.~(\ref{rho_Lambda})
\begin{gather}
\mathcal{A}_{t}=A_{1}+A_{2}-iA_{3}, \quad     \mathcal{B}_{t}=A_{1}-A_{2}-iA_{3}, \quad \mathcal{C}_{t}=2iA_{3},
\end{gather}
\begin{gather}
A_{1} = \frac{m^{2}}{8 \hbar^2t^2 \sigma_0^2 B^2 }, \quad
B^2 =  \frac{1}{4\sigma_0^4} + \frac{1}{2\ell_0^2\sigma_0^2} +\left( \frac{m}{2 \hbar t} + \frac{\gamma}{2\sigma_0^2} \right)^2  + \frac{\Lambda t}{3\sigma_0^2},
\end{gather}
\begin{gather}
A_2 = \frac{m^{2}}{4 \hbar^2t^2 B^2} \left( \frac{1}{2 \ell_0^2} + \Lambda t \right) + \frac{\Lambda t}{12 \sigma_0^2 B^2} \left( \Lambda t + \frac{1}{2\sigma_0^2} + \frac{2}{\ell_0^2} \right) 
+ \frac{m \Lambda \gamma}{4 \hbar \sigma_0^2 B^2} + \frac{\Lambda t \gamma^2}{12 \sigma_0^4 B^2},
\end{gather}
and
\begin{align}
A_3 = \frac{m}{4 \hbar t \sigma_0^2 B^2} \left( \Lambda t + \frac{1}{2\sigma_0^2} + \frac{1}{\ell_0^2} \right) + \frac{m \gamma}{8 \hbar t \sigma_0^2 B^2}\left( \frac{m}{ \hbar t} + \frac{\gamma}{\sigma_0^2} \right),  \;\;\;\;\;\;\; \mathcal{N}_t = \sqrt{\frac{2A_{1}}{\pi}}.
\end{align}

Besides this, employing the density matrix, Eq.~(\ref{rho_Lambda}), and the parameters (C1-C4), the purity Eq. (\ref{eq:purity_new}), after some algebraic simplifications, can be explicitly written as follows 
\begin{gather}
\mu(\gamma,\Lambda,t)=\Bigg[1+\frac{2\sigma_{0}^{2}}{\ell_{0}^{2}}+4\sigma_{0}^{2}\Lambda t+\frac{4\gamma\Lambda\hbar}{m}t^{2}+ 
\frac{4\hbar\Lambda(\gamma^{2}+1+\frac{2\sigma_{0}^{2}}{\ell_{0}^{2}})}{3\tau_{0}m}t^{3}+\frac{4\Lambda^{2}\hbar^{2}}{3m^{2}}t^{4}\Bigg]^{-1/2}.
\label{purity}
\end{gather}

\section{QFI parameters for correlation estimation }\label{Ap:QFIgamma}

 In this appendix, we treat the parameters associated with the function on Eq.(\ref{eq:QFIgamma}), representing the QFI for correlation estimation.
\begin{gather}
    \Phi_{\gamma}(\gamma,t,\Lambda)= \frac{1}{72 \ell_0^2 \tau_0^4} \sum_{k = 0}^{2} \mathcal{C}_k \Lambda^k  , 
\end{gather}
where $\mathcal{C}_k = \mathcal{C}_k (\gamma,t,\Lambda)$, are expressed as follows
\begin{align}
 \mathcal{C}_0 = 9 \ell_0^2 \tau_0^4 \left[ 1 + 2 \left( \frac{\sigma_0}{\ell_0} \right)^2 \right], \;\;\;\; \mathcal{C}_1 = 12 \sigma_0^2 \ell_0^2 \tau_0^2 t^3 \left[    \left(  2 \left( \frac{\sigma_0}{\ell_0} \right)^2 + (\gamma^2 + 1) \right) + 3 \gamma \frac{\tau_0}{t} + 3 \left( \frac{\tau_0}{t} \right)^2 \right],   
\end{align}
and
\begin{align}
\mathcal{C}_2 = 32 \sigma_0^4 \ell_0^2 t^6 \left[ \gamma^2 + 3 \gamma \frac{\tau_0}{t} + \frac{21}{8} \left(\frac{\tau_0}{t} \right)^2 \right].
\end{align}

\begin{figure}[ht]
\centering
\includegraphics[scale=0.36  ]{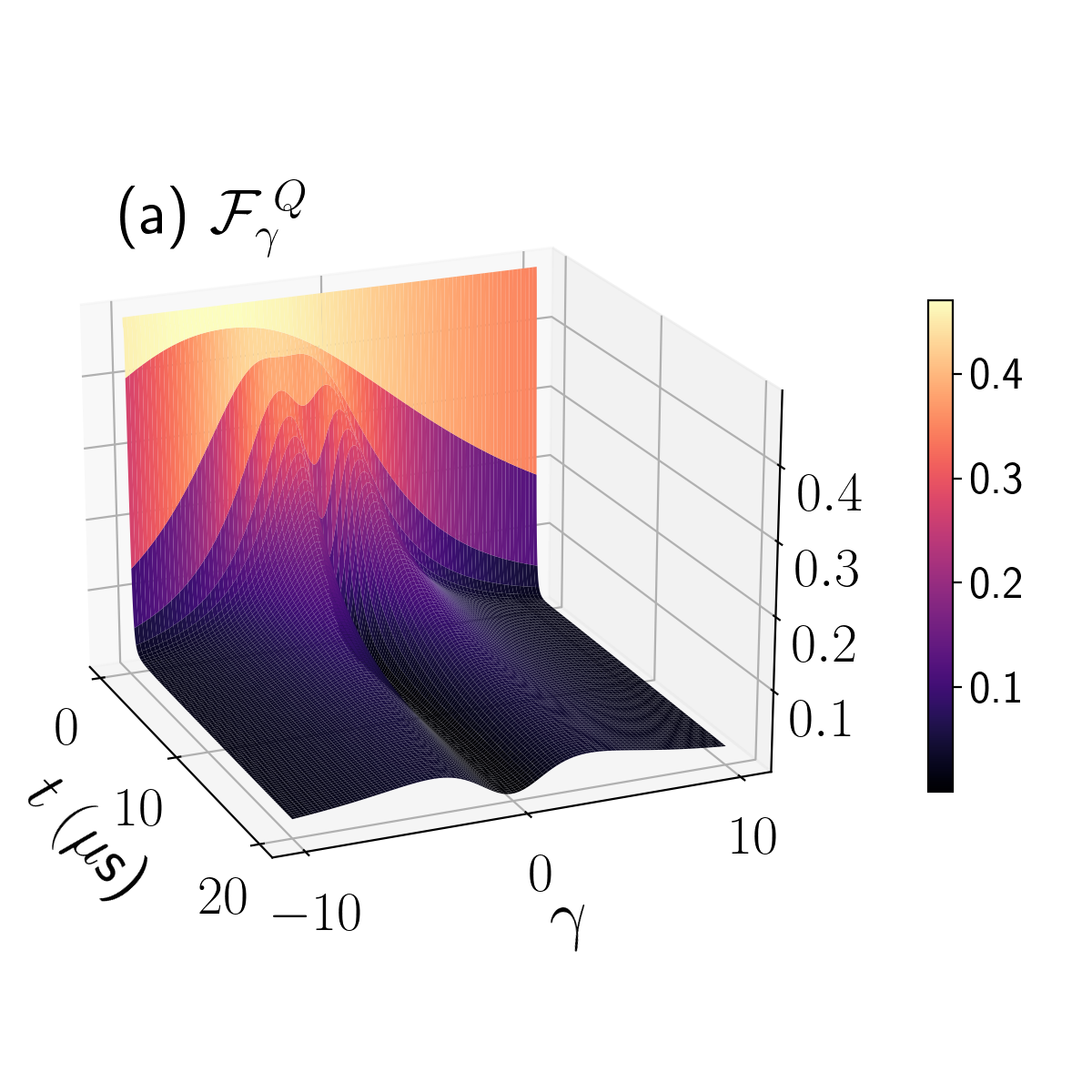}
\includegraphics[scale=0.36]{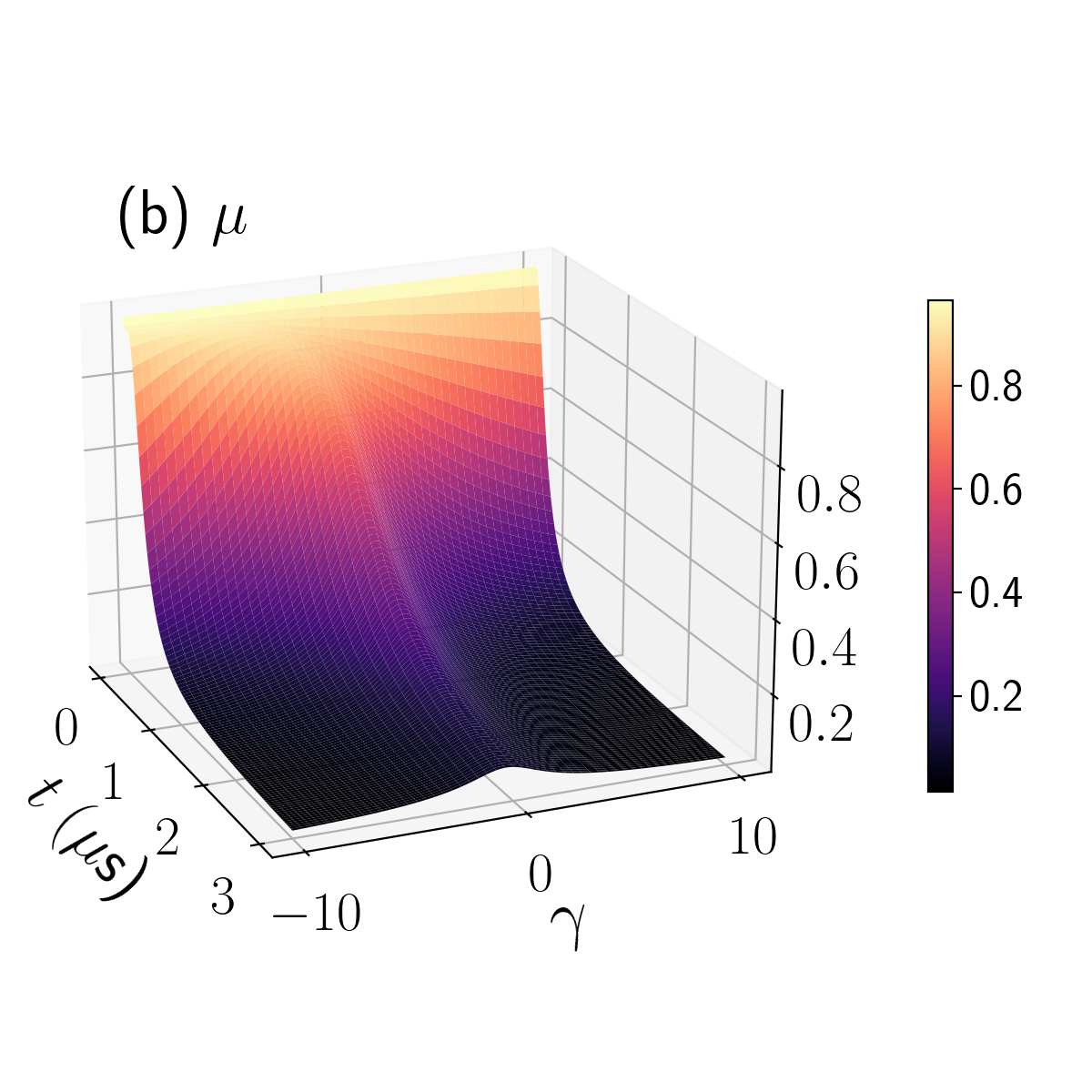}
\includegraphics[scale=0.36]{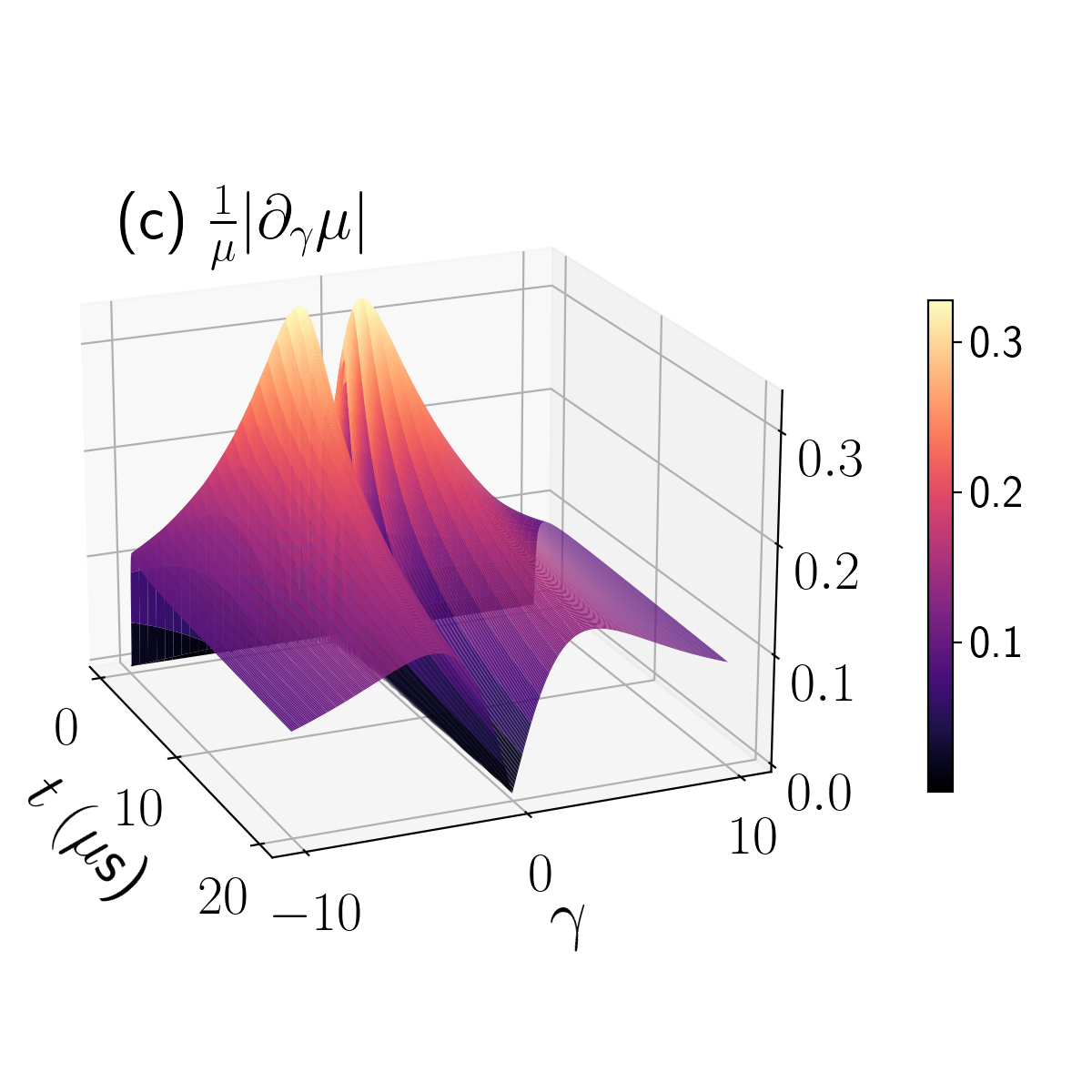}
\caption{(a) QFI ($\mathcal{F}_{\gamma}^{\text{\;Q}}$), (b) purity $\mu$ and (c) its absolute relative derivative $\frac{1}{\mu}|\partial_{\gamma}\mu|$ as a function of initial correlation $\gamma$ and the propagation time $t$, considering $\Lambda = 10^{22}$ m$^{-2}$s$^{-1}$ ($T=20.5 \times 10^3$ K). We can observe that, for small (larger) propagation times $t$, the QFI behavior is governed by the purity  (absolute variation in purity).  }\label{3d}
\end{figure}

We show in Fig. \ref{3d} (a) QFI ($\mathcal{F}_{\gamma}^{\text{\;Q}}$), (b) purity $\mu $ and (c) its absolute relative derivative $\frac{1}{\mu}|\partial_{\gamma}\mu|$ as a function of initial correlation $\gamma$ and the propagation time $t$, considering $\Lambda = 10^{22}$ m$^{-2}$s$^{-1}$ ($T=20.5 \times 10^3$ K). Observe that the QFI behavior is directly influenced by the purity (absolute variation in purity) in the regime of small (larger) propagation times $t$, allowing us to describe how the transition occurs between a regime determined by purity or its rate of variation in the behavior of the QFI.

\section{QFI parameters for the effective environmental coupling estimation }\label{Ap:QFILambda}
 In this appendix, let us show the constants associated with the estimation of the environment coupling through the QFI [$ \mathcal{F}_{\Lambda}^{\text{C}}(\gamma,\Lambda,t)$].
 \begin{figure}[ht]
\centering
\includegraphics[scale=0.25]{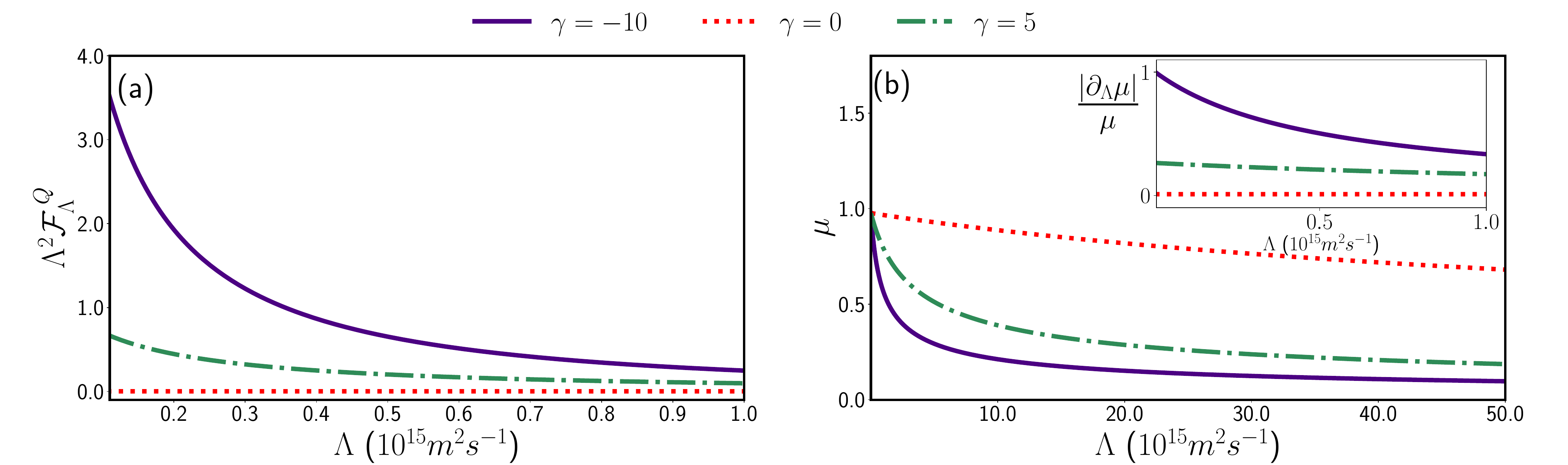}
\caption{(a) QFI ($\Lambda^2\mathcal{F}_{\Lambda}^{\text{\;Q}}$) and (b) purity $\mu$ as a function of environmental effect $\Lambda$ for $t =50.0 $ $\mu$s, and considering the following values for the initial correlation: $\gamma = -10$, $\gamma = 0$ and $\gamma = 5$. The inset in (b) shows the absolute relative variation $\frac{1}{\mu}|\partial_{\Lambda}\mu|$, where we can see how it directly affects the way the QFI behaves. Also, note that initial correlated states with $\gamma \neq 0$ show a more resilient QFI when compared with the standard uncorrelated Gaussian state, however, this does not necessarily result in a more robust purity.}\label{fig7}
\end{figure}
\begin{gather}
    \Phi_{\Lambda}(\gamma,t,\Lambda)= \frac{1}{18 \ell_0^4 \tau_0^4} \sum_{k = 0}^{2} \bar{\mathcal{Z}}_k \Lambda^k,  
\end{gather}
where $\bar{\mathcal{Z}}_k = \bar{\mathcal{Z}}_k (\gamma,t,\Lambda)$, expressed as follows
\begin{align}
 \bar{\mathcal{Z}}_0 = 2 \ell_0^4 \sigma_0^4 t^6 \left[ \Gamma^2 + 6 \gamma \frac{\tau_0}{t} \Gamma + 15 \left(\frac{\tau_0}{t}\right)^2 \left(  \frac{3}{5} \left( \frac{\sigma_0}{\ell_0} \right)^2 + \left(\gamma^2 + \frac{3}{10}\right) \right) + 18 \gamma \left(\frac{\tau_0}{t}\right)^3 + 9 \left(\frac{\tau_0}{t}\right)^4 \right],    
\end{align}
\begin{align}
\bar{\mathcal{Z}}_1 = 4 \sigma_0^6 \ell_0^4 t^7 \left[ 2 \left( \frac{\sigma_0}{\ell_0} \right)^2 + \left(\gamma^2 + 1 \right)  + 3 \gamma \frac{\tau_0}{t} + 3 \left( \frac{\tau_0}{t} \right)^2 \right], \;\;\; \bar{\mathcal{Z}}_2 = 4 \sigma_0^8 \ell_0^4 t^8, \;\; \text{and} \;\;\;\;  \Gamma =  \left(  2 \left( \frac{\sigma_0}{\ell_0} \right)^2 + (\gamma^2 + 1) \right). 
\end{align}

In the Fig. \ref{fig7}, we analyze the dependence of the QFI ($\Lambda^2\mathcal{F}_{\Lambda}^{\;\text{Q}}$) and purity $\mu$ as a function of environmental effect $\Lambda$. The inset in Fig.~\ref{fig7}(b) shows how the absolute relative variation $\frac{1}{\mu}|\partial_{\Lambda}\mu|$ directly affects the way the QFI behaves. Furthermore, note that initial correlated states with $\gamma \neq 0$ have more QFI when compared with the standard uncorrelated Gaussian state. Also, although for the standard Gaussian state ($\gamma = 0$), the purity remains almost unchanged, preserving the value of this quantity does not translate into any gain in QFI. Therefore, the PM-correlations can be employed to enhance the thermal sensitivity of the probe in the regime of small environmental effects (low-temperature regime). Moreover, the range of thermal precision improvement is broad, albeit decreases with temperature growth.

\section{Position-momentum correlations and Squeezing}\label{Ap_PM_squeezing}
Here we discuss the connection between PM correlation and squeezing for pure initial states.
The correlated Gaussian state described by Eq. (\ref{psi_0}) was introduced in Ref.~\cite{DODONOV1980PLA}, where the real parameter $\gamma$ ensures the correlation. 
The uncertainty in position and momentum for this initial state is given by $\sigma_{xx}(0)=\sigma_{0}/\sqrt{2}$ and $\sigma_{pp}(0)=(\sqrt{1+\gamma^{2}})\hbar/\sqrt{2}\sigma_{0}$, whereas their covariance becomes
\begin{gather}
    \sigma_{xp}(0)= \langle \psi_0 |( \hat{x}\hat{p} +\hat{p}\hat{x})/2  | \psi_0 \rangle -\langle \psi_0 |\hat{x} |\psi_0 \rangle \langle \psi_0 |\hat{p} |\psi_0 \rangle  =\hbar\gamma/2.
\end{gather}
Exploring the correlation coefficient between $\hat{x}$ and $\hat{p}$, i.e., $r=\sigma_{xp}/\sqrt{\sigma_{x}\sigma_{p}}\,(-1\leq r\leq1)$, the $\gamma$ parameter turns out to be $\gamma=r/\sqrt{1-r^{2}}\,(-\infty\leq \gamma\leq\infty)$, illustrating the physical meaning of $\gamma$ as a parameter that encodes the initial correlations between $\hat{x}$ and $\hat{p}$. 
Note that, it was explicitly considered the quantized operators $\hat{x}$ and $\hat{p}$, with $[\hat{x},\hat{p}]=i\hbar$. Also, it was employed the symmetrization procedure to transform the product operator $\hat{x}\hat{p}$ into a hermitian operator $(\hat{x}\hat{p} +\hat{p}\hat{x})/2$. 
From a practice point of view, this parameter can be seen as due to an atomic beam propagation along a transverse harmonic potential that effectively acts as a thin lens, which leads to a quadratic phase shift in the initial state \cite{Janicke1995} (see also Appendix \ref{Ap:correlated_states}). Furthermore, Gaussian correlated packets have been used in many contexts, for example in quantum optics \cite{Campos1999JMO}.

The variances in position and momentum for the dimensionless operators $\hat{x}$ and $\hat{p}$ are given by~\cite{MarinhoPRA2020}  
\begin{equation}
    \langle \psi_0 |(\Delta \hat{x})^2  | \psi_0 \rangle = \frac{1}{2};  \quad  \langle \psi_0 |(\Delta \hat{p})^2  | \psi_0 \rangle = \frac{1+\gamma^2}{2}.
\end{equation}
This shows that the correlated Gaussian state is not squeezed in terms of these conventional operators. On the other hand, in terms of the generalized quadratures $\widehat{X1}$ and $\widehat{X2}$, which are defined in terms of  $\hat{x}$ and $\hat{p}$ through a rotation in the phase space by an angle $\theta$ ($\widehat{X1}= \cos\theta \hat{x} + \sin\theta \hat{p}$; $\widehat{X2}= -\sin\theta \hat{x} + \cos\theta \hat{p}$), this state presents
squeezing, as discussed in Ref.~\cite{MarinhoPRA2020}. The variances of the new operators concerning the correlated Gaussian state are
\begin{equation}
    \langle \psi_0 |(\Delta \widehat{X1})^2  | \psi_0 \rangle = \frac{1}{2}[1+\gamma\sin 2\theta + \gamma^2 \sin^2 \theta] \quad  \text{and} \quad \langle \psi_0 |(\Delta \widehat{X2})^2  | \psi_0 \rangle = \frac{1}{2}[1-\gamma\sin 2\theta + \gamma^2 \cos^2 \theta].
\end{equation}
For more details about the connection between PM correlations and squeezing we refer the reader to Ref.~\cite{Marinho2024SciRep}, where the variances of the operators $\widehat{X1}$ and $\widehat{X2}$ are investigated as a function of the rotation angle $\theta$ for the initially correlated Gaussian state.

\bibliography{references}

\end{document}